\documentclass[traditabstract]{aa} 
\usepackage{epsfig,amsmath}
\usepackage[dvips]{color}
\usepackage{graphicx}
\usepackage{txfonts}
\usepackage{natbib}
\bibpunct{(}{)}{;}{a}{}{,}
\usepackage{hyperref}
\usepackage{rotating}

\definecolor{Black}{named}{Black}
\definecolor{Red}{named}{Red}
\definecolor{Green}{named}{Green}
\definecolor{Blue}{named}{Blue}

\begin{document}

\title{Extinction curve template for intrinsically reddened quasars}

\author{T. Zafar\inst{1}
\and P. M{\o}ller\inst{1}
\and D. Watson\inst{2}
\and J. P. U. Fynbo\inst{2}
\and J.-K. Krogager\inst{2}
\and N. Zafar\inst{3}
\and F. G. Saturni\inst{4,1}
\and S. Geier\inst{5,6,7}
\and B. P. Venemans\inst{8}}

\institute{European Southern Observatory, Karl-Schwarzschild-Strasse 2, 85748, Garching, Germany.
 \and Dark Cosmology Centre, Niels Bohr Institute, University of Copenhagen, Juliane Maries Vej 30, DK-2100 Copenhagen, Denmark.
\and Department of Space Science, University of the Punjab, Quaid-i-Azam Campus, Lahore 54590, Pakistan.
\and Dipartimento di Fisica Universit$\grave{\rm a}$ di Roma La Sapienza, Piazzale Aldo Moro 5, 00185 Roma, Italy.
\and Instituto de Astrof\'{i}sica de Canarias, C/ V\'{i}a L$\acute{\rm x}$ctea, s/n, 38205, La Laguna, Tenerife, Spain
\and Departamento de Astrof\'{i}sica, Universidad de La Laguna, 38206 La Laguna, Tenerife, Spain
\and Gran Telescopio Canarias (GRANTECAN), 38205 San Crist$\acute{\rm o}$bal de La Laguna, Tenerife, Spain
\and Max-Planck Institute for Astronomy, K\"{o}nigstuhl 17, D-69117 Heidelberg, Germany}

\titlerunning{UV-steep extinction curves in intrinsically reddened quasars}
\authorrunning{T. Zafar et al.}

\offprints{tzafar@eso.org}

\date{Received  / Accepted }

\abstract {We analyzed the near-infrared to UV data of 16 quasars with redshifts ranging from 0.71 $<$ $z$ $<$ 2.13 to investigate dust extinction properties. The sample presented in this work was obtained from the High $A_V$ Quasar (HAQ) survey. The quasar candidates were selected from the Sloan Digital Sky Survey (SDSS) and the UKIRT Infrared Deep Sky Survey (UKIDSS), and follow-up spectroscopy was carried out at the Nordic Optical Telescope (NOT) and the New Technology Telescope (NTT). To study dust extinction curves intrinsic to the quasars, we selected 16 cases from the HAQ survey for which the Small Magellanic Cloud (SMC) law could not provide a good solution to the spectral energy distributions (SEDs). We derived the extinction curves using the Fitzpatrick \& Massa 1986 (FM) law by comparing the observed SEDs to a combined previously published quasar template. The derived extinction, $A_V$, ranges from 0.2--1.0\,mag. All the individual extinction curves of our quasars are steeper ($R_V=2.2$--2.7) than that of the SMC, with a weighted mean value of $R_V=2.4$. We derived an average quasar extinction curve for our sample by simultaneously fitting SEDs by using the weighted mean values of the FM law parameters and a varying $R_V$. The entire sample is well fit with a single best-fit value of $R_V=2.2\pm0.2$. The average quasar extinction curve deviates from the steepest Milky Way and SMC extinction curves at a confidence level $\gtrsim95\%$. Such steep extinction curves suggest that a significant population of silicates is involved in producing small dust grains. Another possibility might be that the large dust grains may have been destroyed by the activity of the nearby active galactic nuclei (AGN), resulting in steep extinction curves.}

\maketitle{}

\section{Introduction}
The interstellar medium (ISM) is full of small condensed particles called interstellar dust, which play a crucial role in the formation of stellar populations. The dust scatters and absorbs the UV and optical light and affects our measurements of distant objects;
this needs to be corrected for any quantitative analysis. To study dust, the standard method is to determine the extinction curve by fitting an empirical extinction law, thereby revealing information about the total column density of dust grains, their sizes, and compositions. For example, the extinction curves of the Milky Way (MW) and the Large and Small Magellanic Clouds (LMC and SMC) are different from each other \citep{fm86,ccm89,pei92}, which is partly due to the presence and relative strength of the so-called 2175\,\AA\ dust absorption feature, but also to the UV-steepness. 

While the use of empirical extinction laws in most cases provides
a good basis for classifying different types of curves \citep{pei92}, it has proven necessary in some cases to add adjustable parameters to the laws \citep{fm86}. Of particular interest is the so-called total-to-selective extinction $R_V$ (where $R_V=A_V/E(B-V)$), where a low value of $R_V$ corresponds to
a steep curve.  In the classic empirical curves of MW, LMC,
and SMC, $R_V$ has values of 3.08, 3.16, and 2.93, respectively \citep{pei92}.

The Sloan Digital Sky Survey (SDSS) statistical study of the colours of quasi-stellar objects (QSOs) demonstrated the existence of a significant population of dusty QSOs \citep{richards03}. Heavily reddened QSOs have been widely detected \citep[e.g.][]{hall02,ellison04,wang05,jorgenson06,meusinger12,glikman12,fynbo13}. The SMC extinction curve commonly is used to de-redden QSOs \citep[e.g.][]{richards03,hopkins04,glikman12}. The 2175\,\AA\ extinction feature has also been detected in a few QSOs \citep[e.g.][]{jiang11}. A grey extinction curve, flatter than the SMC curve, has been proposed in some cases \citep{maiolino01,gaskell04,czerny04,gaskell07}. The supernova dust extinction curve, which is flat at $\lambda_{\rm rest}$$>$1700\,\AA\ and steeply rising at short wavelengths, has been observed in some $z$ $>$ 4 QSOs \citep{maiolino04,gallerani10}. \citet{hall02} reported two broad absorption line (BAL) QSOs with extinction curves even steeper than the SMC curve. It is important to note that dust in the intervening absorption line system can also contribute to the excess reddening toward QSOs \citep[e.g.][]{menard08,wang12}, and care should be taken to
avoid confusing the two.

In a recent survey for High $A_V$ QSOs (HAQ, \citealt{fynbo13}) it was found that a significant fraction of the highly reddened QSOs have very steep reddening curves that cannot be matched by the usual SMC extinction curve. Here we aim to characterize this sub-class of QSOs in more
detail and in particular to determine whether they are similar enough
to allow a global description. The paper is organized as follows: in Sect. 2 we present our data and sample selection criteria. In Sect. 3 we define the dust model, and Sect. 4 describe results from the analysis. Section 5 provides a discussion, and Sect.
6 summarizes conclusions.

\section{Data and sample definition}
\subsection{Multi-wavelength data}
The QSO data used in this study is a sub-sample of the HAQ survey from \citet[][hereafter paper\,I]{fynbo13} and \citet[][hereafter paper\,II]{krogager14}. The paper\,II survey is an extension of paper\,I where QSOs were selected from the SDSS and the United Kingdom InfraRed Telescope (UKIRT) Infrared Deep Sky Survey (UKIDSS) colours. The near-infrared (NIR) to UV data for these quasars are completed by taking advantage of the SDSS ($u$, $g$, $r$, $i$, and $z$) together with UKIDSS ($Y$, $J$, $H$, $K_s$) catalogues \citep{peth11}. The details on building the sample are presented in papers\,I and\ II. Follow-up spectroscopy confirmed that 79\% (paper\,I) and 97\% (paper\,II) candidates are QSOs. The spectra were taken with the Nordic Optical Telescope (NOT) equipped with the instrument Andalucia Faint Object Spectrograph and Camera (ALFOSC) and the New Technology Telescope (NTT) equipped with the ESO Faint Object Spectrograph and Camera 2 (EFOSC2). The spectra were then processed and flux calibrated using standard data reduction techniques within IRAF and MIDAS. Both spectra and photometry were corrected for Galactic extinction using the dust maps from \citet{schlegel98}. All the optical spectra were scaled to the $r$-band photometry to fix their absolute flux calibration to that of the photometry. The spectra were not corrected for telluric absorption. 

\subsection{Sample definition}
Papers\,I and\ II advocated that in most cases dust is located at the redshift of the QSO. Paper\,I found no evidence of intervening absorbing galaxies that could be responsible for the reddening. However, paper\,II found 9 out of 154 QSOs for which the model with dust from the intervening absorber is preferred. Papers\,I and\ II found that for the majority of the QSOs, the QSO template reddened by the SMC-type extinction curve provides a good match to the data. However, for some QSOs, the same simplistic model fails in the NIR. Paper\,I further argued that this might be due to $i)$ a problem with the QSO template, most likely containing significant contamination from the host galaxy, or $ii)$ usage of a simplistic SMC extinction curve for these data, where the latter is more probable. Instead of using a simplistic SMC model, we here use an elaborate dust law to fit these excessive UV-flux-deficient cases. 

To study extinction curves intrinsic to the QSOs, we considered as our basic sample all QSOs from papers\,I and\ II, excluding the nine objects for which the original analysis indicated that the absorption might be caused by an intervening absorber. We
combined the remaining objects into a sub-sample of 31 QSOs where the QSO composite spectrum reddened by an SMC extinction curve does not provide a good solution for the QSO spectral energy distribution (SED) and the SMC-reddened QSO template deviates by more than $3\sigma$ from the observed SED. Fifteen of these show a sudden UV-break in the spectra, and a single steep curve cannot fit those data. A detailed analysis of these UV-break QSOs is underway (Zafar et al. 2015, in prep.). The remaining 16 QSOs make up the sample for the present study and have redshifts ranging from $0.71 < z < 2.13$. The details of each QSO are provided in Appendix \ref{ind_qso}. 

\begin{table*}
\begin{minipage}[t]{\columnwidth}
\caption{FM best-fit parameters for the QSO sample.}      
\label{best-fit} 
\centering
\renewcommand{\footnoterule}{}  
\setlength{\tabcolsep}{4pt}
\begin{tabular}{l c c c c c c c}\hline\hline                       
QSO & $c_1$ & $c_2$ & $c_4$ & $c_5$ & $R_V$ & $A_V$ & $\chi^2_{\nu}$ \\
        &  & $\mu$m &  $\mu$m$^2$ & $\mu$m$^{-1}$       & & mag\\ 
\hline
HAQ\,J001522.0$+$112959 & $-4.334\pm0.367$ & $2.521\pm0.310$ & $0.469\pm0.232$ & $5.90\pm0.18$ & $2.66\pm0.31$ & $0.58\pm0.13$ & 1.19 \\
HAQ\,J012925.8$-$005900 & $-4.459\pm0.310$ & $2.406\pm0.287$ & $0.874\pm0.142$ & $5.84\pm0.17$ & $2.41\pm0.28$ & $0.74\pm0.17$ & 1.01 \\
HAQ\,J013016.5$+$143953 & $-4.435\pm0.437$ & $2.371\pm0.370$ & $0.861\pm0.175$ & $5.86\pm0.19$ & $2.23\pm0.35$ & $0.46\pm0.12$ & 0.90 \\
HAQ\,J015136.7$+$061831 & $-4.681\pm0.429$ & $2.421\pm0.456$ & $0.421\pm0.143$ & $5.89\pm0.16$ & $2.29\pm0.23$ & $0.51\pm0.14$ & 1.04 \\
HAQ\,J024717.3$-$005205 & $-4.623\pm0.321$ & $2.326\pm0.205$ & $0.474\pm0.138$ & $5.90\pm0.17$ & $2.52\pm0.19$ & $1.00\pm0.11$ & 1.10 \\
HAQ\,J031213.4$+$003554 & $-4.683\pm0.297$ & $2.365\pm0.232$ & $0.531\pm0.152$ & $5.87\pm0.16$ & $2.29\pm0.23$ & $0.50\pm0.12$ & 1.08 \\
HAQ\,J031901.8$+$062339 & $-5.652\pm0.420$ & $2.454\pm0.413$ & $0.948\pm0.208$ & $5.78\pm0.30$ & $2.38\pm0.35$ & $0.20\pm0.06$ & 0.96 \\
HAQ\,J034748.1$+$011544 & $-4.371\pm0.437$ & $2.369\pm0.394$ & $1.251\pm0.182$ & $6.50\pm0.21$ & $2.23\pm0.39$ & $0.29\pm0.12$ & 1.23 \\
HAQ\,J140047.1$+$021934 & $-4.523\pm0.350$ & $2.316\pm0.268$ & $0.595\pm0.114$ & $5.91\pm0.08$ & $2.16\pm0.36$ & $0.56\pm0.09$ & 1.02 \\
HAQ\,J140952.6$+$094023 & $-4.594\pm0.616$ & $2.398\pm0.343$ & $0.861\pm0.163$ & $5.84\pm0.09$ & $2.27\pm0.26$ & $0.54\pm0.10$ & 1.21 \\
HAQ\,J152710.9$+$025019 & $-5.413\pm0.438$ & $2.241\pm0.289$ & $0.621\pm0.259$ & $5.84\pm0.22$ & $2.33\pm0.36$ & $0.21\pm0.07$ & 1.04 \\
HAQ\,J160601.1$+$290218 & $-4.833\pm0.486$ & $2.361\pm0.310$ & $0.416\pm0.231$ & $5.80\pm0.17$ & $2.44\pm0.24$ & $0.26\pm0.08$ & 1.03 \\
HAQ\,J163957.9$+$315726 & $-4.659\pm0.410$ & $2.344\pm0.194$ & $0.414\pm0.125$ & $6.02\pm0.11$ & $2.48\pm0.18$ & $0.73\pm0.11$ & 1.04 \\
HAQ\,J164332.8$+$294423 & $-4.462\pm0.491$ & $2.323\pm0.258$ & $0.421\pm0.153$ & $5.85\pm0.10$ & $2.56\pm0.27$ & $0.39\pm0.13$ & 1.29 \\
HAQ\,J231046.9$+$111721 & $-4.663\pm0.613$ & $2.364\pm0.482$ & $0.461\pm0.207$ & $5.92\pm0.15$ & $2.44\pm0.28$ & $0.57\pm0.13$ & 1.17 \\
HAQ\,J235526.8$-$004154 & $-4.785\pm0.402$ & $2.281\pm0.325$ & $0.628\pm0.118$ & $5.89\pm0.17$ & $2.54\pm0.24$ & $0.65\pm0.13$ & 1.09 \\
\hline
weighted mean values    & $-4.678\pm0.090$ & $2.355\pm0.017$ & $0.622\pm0.062$ & $5.90\pm0.04$ & $2.41\pm0.04$ & $\cdots$ & $\cdots$ \\
\hline
\end{tabular}
\end{minipage}
\end{table*}

\begin{figure}
  \centering
{\includegraphics[width=\columnwidth,clip=]{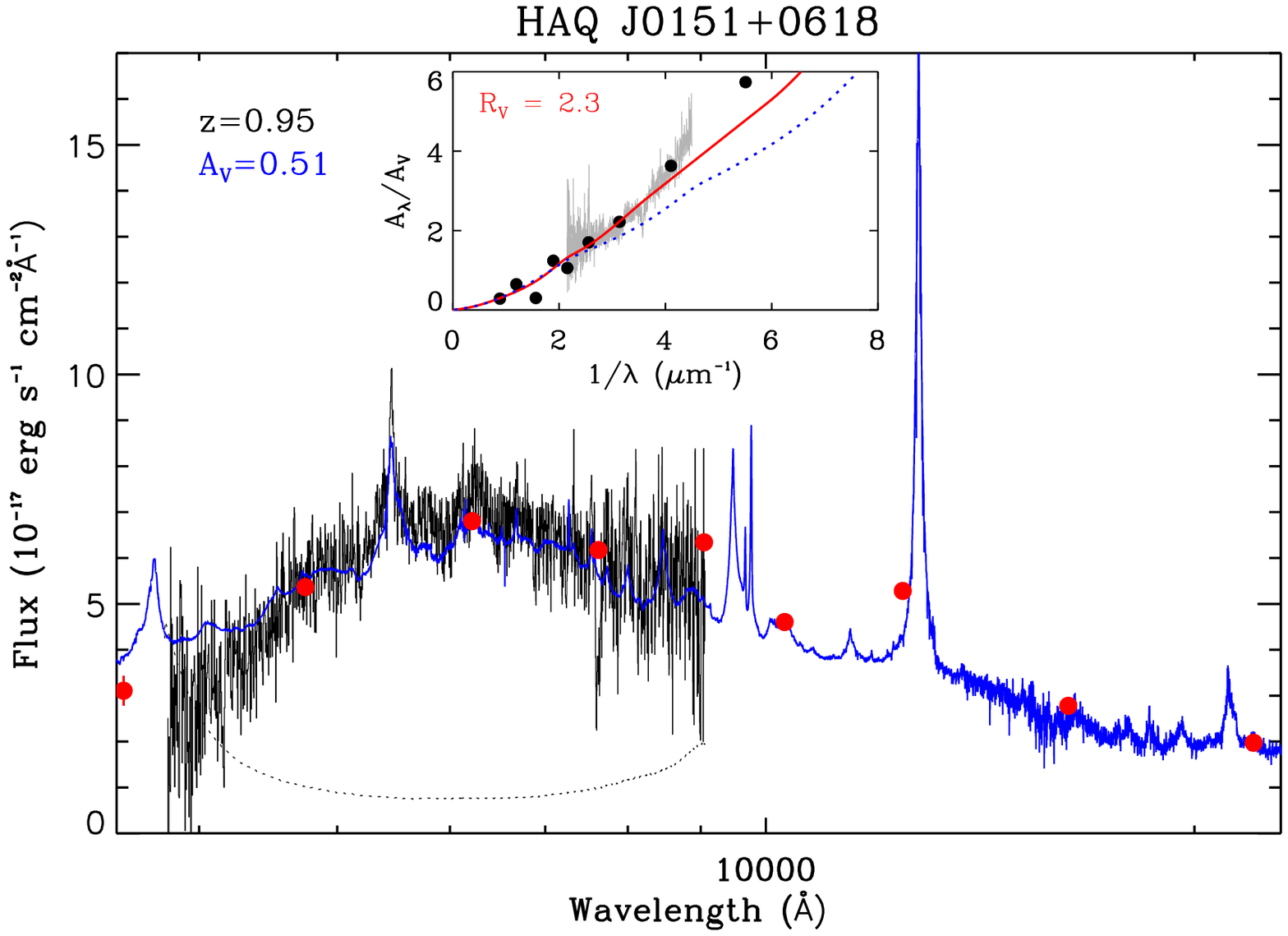}}
     \caption{SED of HAQ\,J0151$+$0618. The solid and dotted lines represent the observed and  error spectrum, respectively. The spectra are binned by a factor of 2 for plotting purposes. The redshift and the rest-frame visual extinction, $A_V$, is provided in the top left corner. The blue curve represents the combined QSO template from \citet{vanden01} and \citet{gilkman06} redshifted to the estimated redshift and reddened by the indicated amount of extinction. Red filled circles correspond to the SDSS and UKIDSS photometry. 
{\it Inset}: Extinction curve of HAQ\,J0151+0618 based on the best-fit model given in Table \ref{best-fit}. The grey curve represents the spectrum. The black circles correspond to the SDSS and UKIDSS photometry. 
The red solid curve corresponds to the best-fit dust extinction model. The average SMC Bar (blue dotted line) extinction model from \citet{gordon03} is also shown. {\it The complete SEDs of the QSO sample are shown in \ref{qso:fits}}}
         \label{q0151}
           \end{figure}

\section{Dust modelling}
\subsection{Fitting procedure}
We used the combined QSO template from \citet{vanden01} and \citet{gilkman06} as a reference for the intrinsic slope of the SEDs of our QSOs (for details see paper\,I). For each QSO SED, the QSO template was redshifted to the redshift of the observed QSO. The QSO template was then normalized to the observed $K_s$-band photometry. This was done because the $K_s$-band is the furthest band available in these SEDs and usually occurs in the rest-frame optical/NIR of the QSOs and is less affected by dust compared to the remaining data. Our spectra fully cover the range of the $g$, $r$, and $i$ photometric points and contain more information. We therefore did not include those three points in the fitting procedure. The spectrophotometric data on the blue side of the QSO Ly$\alpha$ emission are in the Ly$\alpha$ forest and were therefore excluded. We did not consider photometric upper limits.

The QSO-SEDs were fitted with the extinction law of \citet{fm86}, and the best-fit parameters were derived. The dust law is briefly described in Sect. \ref{dust_law}. 
The QSO SEDs were modelled in IDL using the $\chi^2$ minimization algorithm. To obtain errors on each parameter, we used 1000 Monte Carlo (MC) realizations of the data. The mean value of each datapoint was set to the observed value, and a Gaussian distribution was defined with a width corresponding to the 1$\sigma$ measured error on that data point. We then fit the extinction law and estimated the best-fit parameters for each simulated set. For each parameter, the error was calculated using the standard deviation of its distribution. The number of degrees of freedom was calculated by taking the number of pixels in the spectra plus the photometric points outside the spectral range minus the number of fitting parameters.

\subsection{Dust law}\label{dust_law}
We used the \citet{fm86} law, which provides freedom in reproducing the extinction curves through a set of nine parameters. It contains two components: $i)$ a UV linear component specified by the parameters $c_1$ (intercept) and $c_2$ (slope), and parameters $c_4$ and $c_5$  provide the far-UV curvature and $ii)$ a Drude component describing the 2175\,\AA\ bump by the parameters $c_3$ (bump strength), $x_0$ (central wave number), and $\gamma$ (width of the bump). The extinction properties in the NIR and optical were determined using spline interpolation points. For more details of the law see \citet{zafar11,zafar12}. Hereafter we refer to this extinction law as FM. 

\section{Results}
\subsection{Is there a bump at 2175\,\AA?}
As a first step. we fit the QSO SEDs individually using all the nine dust law parameters, but parameter  $c_3$ did not vary significantly
from zero in any of the cases. This suggests that there is no
bump at 2175\,\AA\ in our objects. To test this, we then froze the bump to the level reported by \citet{gordon03} for the average SMC Bar sample and computed the best-fit $\chi^2$s by varying the remaining six parameters. The three fixed values are $c_3=0.389$, $x_0=4.60$ ($\mu$m$^{-1}$), and $\gamma=1.0$ ($\mu$m$^{-1}$). We then performed the individual SED fitting, but now froze $c_3$ to zero, which decreased the $\chi^2_{\nu}$ for all our 16 QSOs from $\chi^2_{\nu}=$ 1.12 to 1.10 for the same numbers of degrees of freedom.
Based on these tests, we conclude that there is no evidence for a bump at 2175\,\AA\  in any of our SEDs. We therefore use $c_3=0$ henceforth. The best-fit extinction curve parameters, with $c_3=0$, of our QSO sample are provided in Table \ref{best-fit}. 

\subsection{Total-to-selective extinction, $R_V$}
\begin{figure}
  \centering
{\includegraphics[width=\columnwidth,clip=]{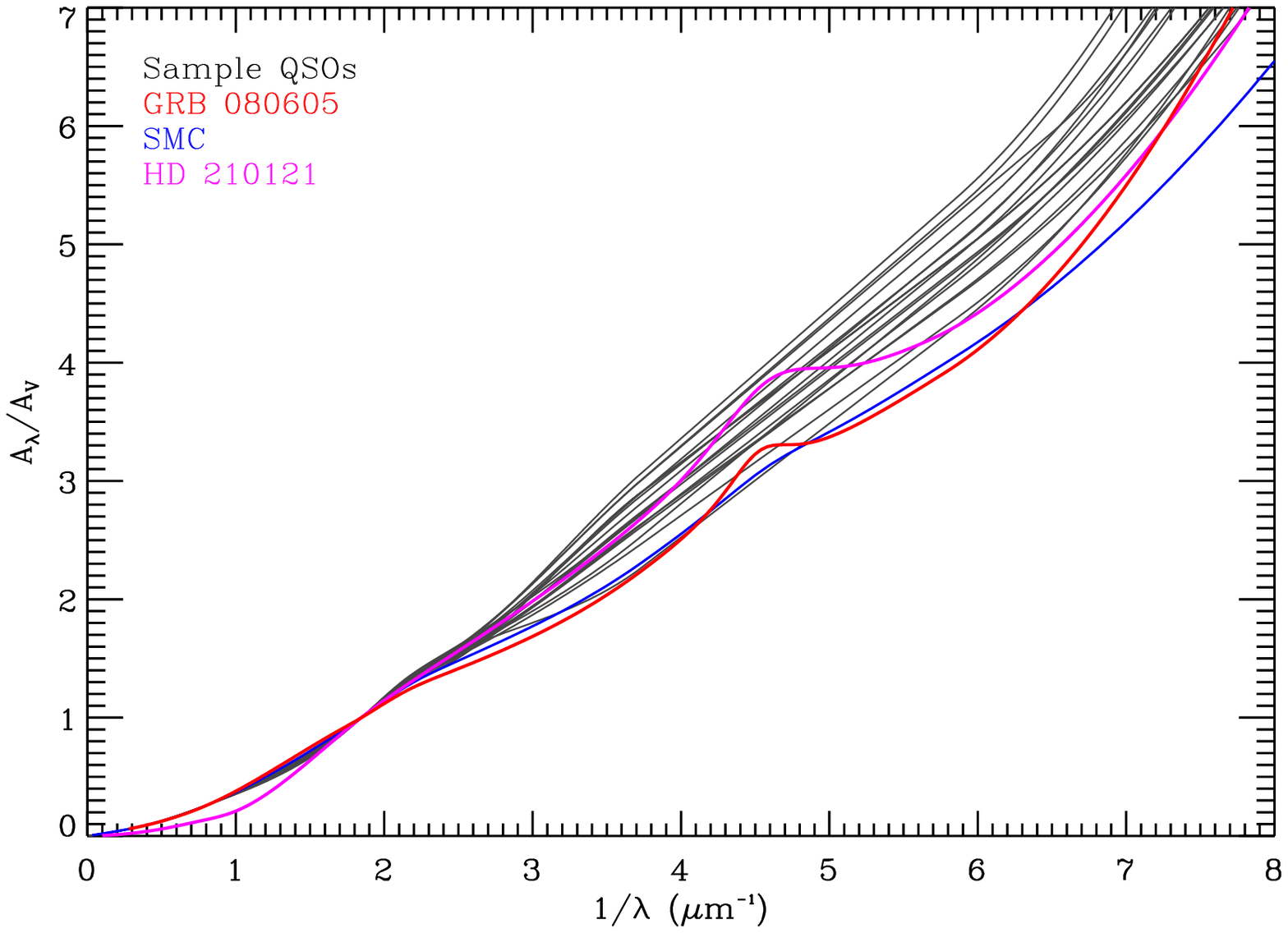}}
\caption{Individual extinction curves of the QSO sample compared with the extinction curves of GRB\,080605 \citep{zafar12}, average SMC Bar \citep{gordon03}, and MW sightline HD\,210121 \citep{fm07}. The QSO extinction curves derived in this work appear to be steeper than the other environments.}
         \label{fig:mean}
\end{figure}

We determined $A_V$s for each of our QSO by fitting the individual SEDs. Our sample fit the steep extinction curves well and has moderate to high extinction values with $A_V$ ranging from $0.20\pm0.06$ to $1.00\pm0.11$. The mean $A_V$ of our sample is $\langle A_V \rangle=0.51$ with a standard deviation of 0.21. In contrast, Table \ref{best-fit} shows that the total-to-selective
extinction ($R_V$) values all are identical to within 1$\sigma$, that is, the sample is consistent with having a single value for $R_V$.  The optimal weighted combined value of individually determined $R_V$s is $2.41\pm0.04,$ which is steeper (Fig. \ref{fig:mean}) than the average SMC Bar \citep[$R_V$=$2.74\pm0.13$;][]{gordon03}.

\subsection{Average QSO extinction curve}
Encouraged by our finding in the previous section that suggested that
$R_V$ is identical for all our QSOs, we now tested whether a single unique
extinction curve might fit our entire sample. First we computed weighted
averages of $c_1$, $c_2$, $c_4$, $c_5,$ and $R_V$ using inverse variance
optimal weights. We report them at the end of Table \ref{best-fit}. To test whether this single extinction curve fit all 16 objects, we then kept all those five parameters fixed and now fit each
object only for $A_V$. The combined $\chi^2$ per degree of freedom ($\chi^2_{\nu}$)
is found to be 1.07, which represents an insignificant change from
1.09, which was the combined $\chi^2_{\nu}$ of the
individual fits.

As a last step we now kept only the four $c$-parameters fixed, while
we performed a simultaneous fit to all 16 objects listed in Table \ref{best-fit}. For this fit we searched for
the global minimum of the $\chi^2$ for a single value of $R_V$ while
at the same time fitting for an individual value of $A_V$ for each
QSO. This final fit provides a final optimal value of
$R_V=2.21\pm0.22$ and a total $\chi^2_{\nu}=1.01$. We adopted this as
the global $R_V$ for our sample of steep extinction curve QSOs, which
together with the values of the $c$-parameters listed in
Table \ref{best-fit} provides a steep extinction curve template.
The parametric average QSO extinction curve is shown in
Fig.~\ref{allextfm} and is defined as 
\begin{equation}
\frac{A_{\lambda}}{A_V} = \frac{1}{2.21}\left(-4.678 + 2.355x + 0.622(x-5.90)^2\right) + 1  
,\end{equation}
where $x=\lambda^{-1}$.

\subsection{Quasar spectral slope}
In the previous sections we have consistently used the standard quasar
template described in Sect. 3.1. This template has an underlying
(unreddened) spectral slope of $\beta_{\nu} = -0.5$. It has been shown
\citep{richards03,krawczyk15} 
that actual quasar spectra follow a distribution of spectral slopes
with an rms $( \beta_\nu )=-0.2.$ Here we examine whether this
scatter causes a significant uncertainty in our determination of
$R_V$.

For this purpose we modified the standard template by altering its
slope with $\pm0.2$, that is, we used $\beta_{\nu} = -0.3$ and $-0.7$ in addition
to $\beta_{\nu} = -0.5$. For the test we then selected the quasar with the
lowest value of $R_V$ in Table \ref{best-fit} (HAQ\,J1400$+$0219), the quasar with the highest
(HAQ\,J0015$+$1129), and one with an $R_V$ close to the final mean
(HAQ\,J1606$+$2902). For those three we repeated the exact same fit as described
in Sect. 4.2, except that we now used templates with $\beta_{\nu} = -0.3$ and
$-0.7$.

  \begin{figure}
  \centering
{\includegraphics[width=\columnwidth,clip=]{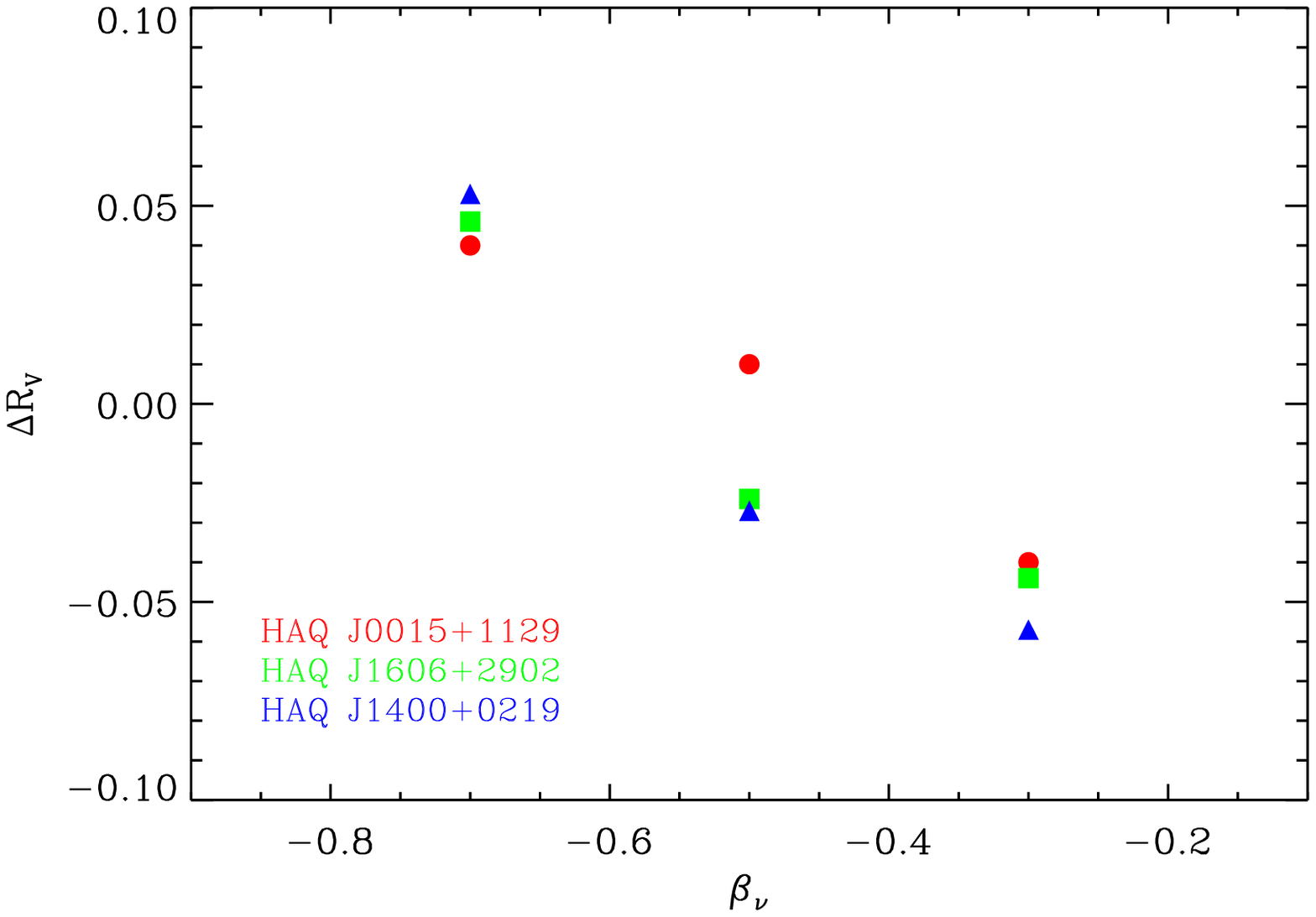}}
     \caption{Change in $R_V$ plotted against quasar intrinsic slope, $\beta_{\nu}$. The three selected quasars HAQ\,J0015$+$1129, HAQ\,J1606$+$2902, and HAQ\,J1400$+$0219 are illustrated in red, green, and blue, respectively.}
         \label{slope}
  \end{figure} 

In Fig. \ref{slope} we plot the change in $R_V$ as a function of $\beta_{\nu}$. Regardless of the value of $R_V$ itself, the added scatter of
$R_V$ is 0.05 for a scatter of 0.2 on $\beta_{\nu}$. Under the realistic assumption,
therefore, that the $\beta_{\nu}$s of our quasar sample follow a
random distribution with scatter 0.2, the final combined additional
error on the combined $R_V$ is
$0.05 / (N)^{0.5} = 0.05 \times 16^{-0.5} = 0.0125,$ which is negligible
considering our final value of $R_V =2.21 \pm 0.22$. We conclude this
section by stating that the combined fit value of $R_V$ is not influenced by
the choice of spectral slope of the quasar template.

\section{Discussion}
The extinction curves and $R_V$ were determined for a wide variety of environments and were found to show considerable dependence on the environment: lower-density regions have a smaller $R_V$ and a steeper far-UV rise ($\lambda^{-1} > 4$\,$\mu$m$^{-1}$), implying formation of smaller dust grains in these environments. Denser regions have a larger $R_V$ and a flatter far-UV curvature, implying larger dust grains \citep{ccm89}. Moreover, dust grains can potentially be affected by the local environments. Strong shocks and UV photons may conspire to destroy large grains and thus change the shape of the extinction curve \citep{jones04}. Elimination of large dust grains could lead to a steepening of the dust extinction curves.

\subsection{Dust composition}
Our QSOs extinction curves are featureless and have a steep rise with $\lambda^{-1}$ into the UV, resembling the SMC curve, however, the additional UV-rise makes these curves peculiar. The absence of the 2175\,\AA\ feature in these extinction curves implies the absence of small carbonaceous grains. There may be small amounts of carbonaceous grains present in these environments, but the steep curve possibly dilutes the strength of the bump. Moreover, the complexity of the QSO spectra prevented us from  quantifying the smaller bumps. The linear steep rise suggests that dust in these environments is composed of smaller grains than that in the Galactic diffuse ISM. This could either be due to more efficient dust destruction as a result of the harsh active galactic nuclei (AGN) environment, or to the low metallicities
of these environments. The former phenomenon is more probable. Moreover, \citet{reach00} also suggested that in hot, luminous environments, very small grains may have been destroyed. \citet{mathis96} proposed that small silicate grains are responsible for the steep UV extinction. \citet{weingartner01} and \citet{li01} developed a grain model consisting of carbonaceous grains producing the
bump at 2175\,\AA\  and the amorphous silicate population that
defines the steep UV curvature. The dust properties observed here may simply be a reflection of such a silicate population.

Silicates are the most commonly found dust species in the ISM of galaxies. Silicate dust makes up $\sim$70\% of the core mass of interstellar dust grains \citep[e.g.][]{draine03}. It is worth noting that 10$\mu$m silicate emission features are discovered in luminous dusty QSOs \citep[e.g.][]{siebenmorgen05,hao05,sturm06}. \citet{kulkarni07,kulkarni11} reported 9.7$\mu$m interstellar silicate absorption feature in QSO absorption systems. Recently, \citet{hatziminaoglou15} have performed a comprehensive study of 9.7$\mu$m and 18$\mu$m silicate emission features in a sample of AGNs. In contrast, observational evidence of local Seyfert galaxies dust tori suggests that large graphite grains dominate the sublimation zone of the inner torus \citep{kishimoto07,kishimoto09,mor09,kishimoto11,mor12}. These studies indicate an altered distribution of grain sizes and their composition.

  \begin{figure}
  \centering
{\includegraphics[width=\columnwidth,clip=]{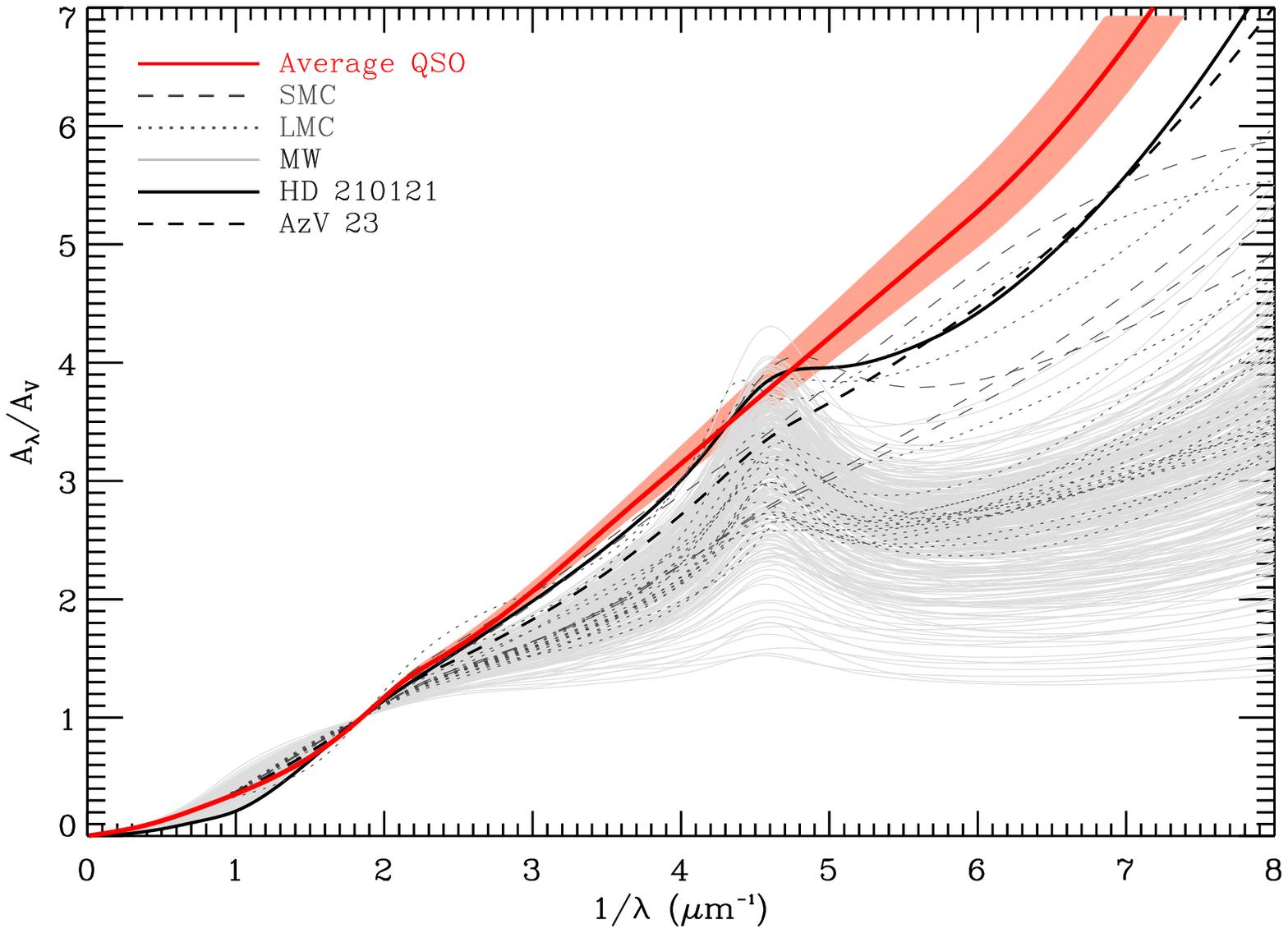}}
     \caption{Average QSO extinction curve of our sample compared with the Local Group extinction curves. The red shaded area illustrates the dispersion in the average QSO extinction curve. The extinction curves of the MW sightlines taken from \citet{fm07} are shown in grey. The SMC and LMC extinction curves taken from \citet{gordon03} are shown in charcoal dashed and dotted lines, respectively. The black solid line represents the extinction curve towards the MW star HD\,210121. The black dashed line corresponds to the extinction curve towards the SMC AzV\,23 sightline.}
         \label{allextfm}
  \end{figure} 

\subsection{Comparison of steep extinction curves}
We also compared our QSO sample results to the MW \citep{fm07}, LMC and SMC \citep{gordon03} sightlines. Most of the extinction curves studied in the Magellanic Clouds are from the active star-forming regions, where strong shocks and UV photons may destroy large dust grains, whereas in the MW the sightlines typically point
towards main-sequence OB stars. The UV rise in the average QSO extinction curve is typically steeper than for the vast majority of known Local Group sightlines for a given value of $A_V$ (Fig. \ref{allextfm}). The steepest SMC extinction curve is seen towards the sightline AzV\,23 with $R_V=2.65$ (\citealt{gordon03}; see Fig.~\ref{allextfm}). Another example of a steep extinction curve with a bump at 2175\,\AA\  is seen in the MW towards HD\,210121 with $R_V=2.4$ \citep{larson00,sofia05}.
However, the average QSO extinction curve deviates from both of these extinction curves at $\gtrsim$95\% confidence level. 

Outside the Local Group, steeper extinction laws are also seen. Extinction curves steeper than the MW with $R_V=2.4$--2.5 and a bump at 2175\,\AA\  are detected in the central M31 bulge \citep{dong14}. \citet{amanullah14} found a steep extinction curve with $R_V=1.4$ for a Type\,Ia supernova SN\,2014JA in the starburst galaxy M82. A sample of $\sim$2000 UV-deficient narrow line Seyfert galaxies also shows a steeper extinction law \citep{zhou06}. Anomalously steep reddening laws have been seen in Seyfert galaxies NGC\,3227 and Mrk\,231 \citep{crenshaw01,leighly14}. \citet{schady12} found UV-steep extinction curves for moderately extinguished Gamma-ray burst (GRB) afterglows and flatter extinction curves for some heavily extinguished GRBs ($A_V>1$). \citet{fynbo14} reported that the observed spectrum of GRB\,140506A can be defined with a steep UV extinction curve and a giant bump at 2175\,\AA. An extinction curve with a steep UV-slope and a flatter $R_V=3.19^{+0.86}_{-0.89}$ with a 2175\,\AA\ feature is found in GRB\,080605 \citep{zafar12};
it deviates from the average QSO extinction curve at $\gtrsim$90\% confidence level. The FM law parameters $c_4$ and $c_5$ define the far-UV curvature. Except for three cases (HAQ\,J0319$+$0623, HAQ\,J1527$+$0250, and HAQ\,J1606$+$2902), we usually lack data below $\lambda_{\rm rest}<1700$\,\AA\ (i.e. $1/\lambda>6$\,$\mu$m$^{-1}$), therefore, we are not able to fully constrain the far-UV curvature. However, for these three cases we do not find any significant UV-turnover.

Previously, \citet{gallerani10} claimed extinction curves deviating from the SMC law and flattening at $\lambda_{\rm rest}$$<$2000\,\AA\ in a sample of QSOs at $3.9<z<6.4$. \citet{gallerani10} selected bright blue-slope QSOs visible during their observing runs and their shallower extinction curves probably driven by the assumed template with an adjustable intrinsic slopes for the QSO continua. Recently, \citet{hjorth13} found that the median extinction curve of QSOs at $z\sim6$ is consistent with the SMC curve. Because of their bright optical magnitudes and blue colours, the QSO
selection naturally favours objects with relatively flat extinction curves. The HAQ survey of papers\,I and\ II explicitly searches for red objects, therefore, we find steep extinction curves in our sample. 

The results of this work imply that the common usage of the SMC extinction law to fit the QSO SEDs is inadequate \citep[see also][]{clayton00,gordon03}. In reality, Local Group sightlines exhibit a variety of extinction curves. This work shows that similar to the Local Group, QSOs without a bump at 2175\,\AA\  seem to have a continuum of steep extinction curves. 

\section{Conclusions}
We presented a UV-flux-deficient sub-sample of the HAQ survey from papers\,I and\ II to study the dust properties of the intrinsically reddened QSOs. We analyzed the NIR to UV SEDs of 16 QSOs with redshifts ranging from 0.71 $<$ $z$ $<$ 2.13. We modelled the rest-frame QSO-SEDs by comparing them to the combined QSO template from \citet{vanden01} and \citet{gilkman06}. These dusty QSOs have $A_V$ values ranging from $A_V=0.2$--1.0\,mag. We found no evidence for a 2175\,\AA\ dust feature. All our QSO-SEDs require an extinction law steeper than that of the usual SMC. We derived the average QSO extinction curve by simultaneously fitting SEDs by fixing the FM law to the combined weighted mean values and fitting for $R_V$ and $A_V$. The entire sample is well fit with a best-fit value of $R_V=2.21\pm0.22$. The average QSO extinction curve deviates from the steepest MW and SMC extinction curves at $\gtrsim$95\% confidence level. Such steep extinction curves may indicate that a significant population of silicates produces small dust grains in the harsh QSO environments. Another possibility is that large dust grains may have been destroyed or cracked by the activity of the nearby AGN, resulting in steep extinction curves. However, an altered distribution of grain sizes is possible.

\begin{acknowledgements}
JPUF acknowledges support from the ERC-StG grant EGGS-278202. The Dark Cosmology Centre is
funded by the DNRF. 
\end{acknowledgements}

\bibliographystyle{aa}
\bibliography{qso_steep_ext.bib}{}

\begin{appendix}
\section{Notes on individual QSOs}\label{ind_qso}
\begin{figure*}
  \centering
{\includegraphics[width=0.63\columnwidth,clip=]{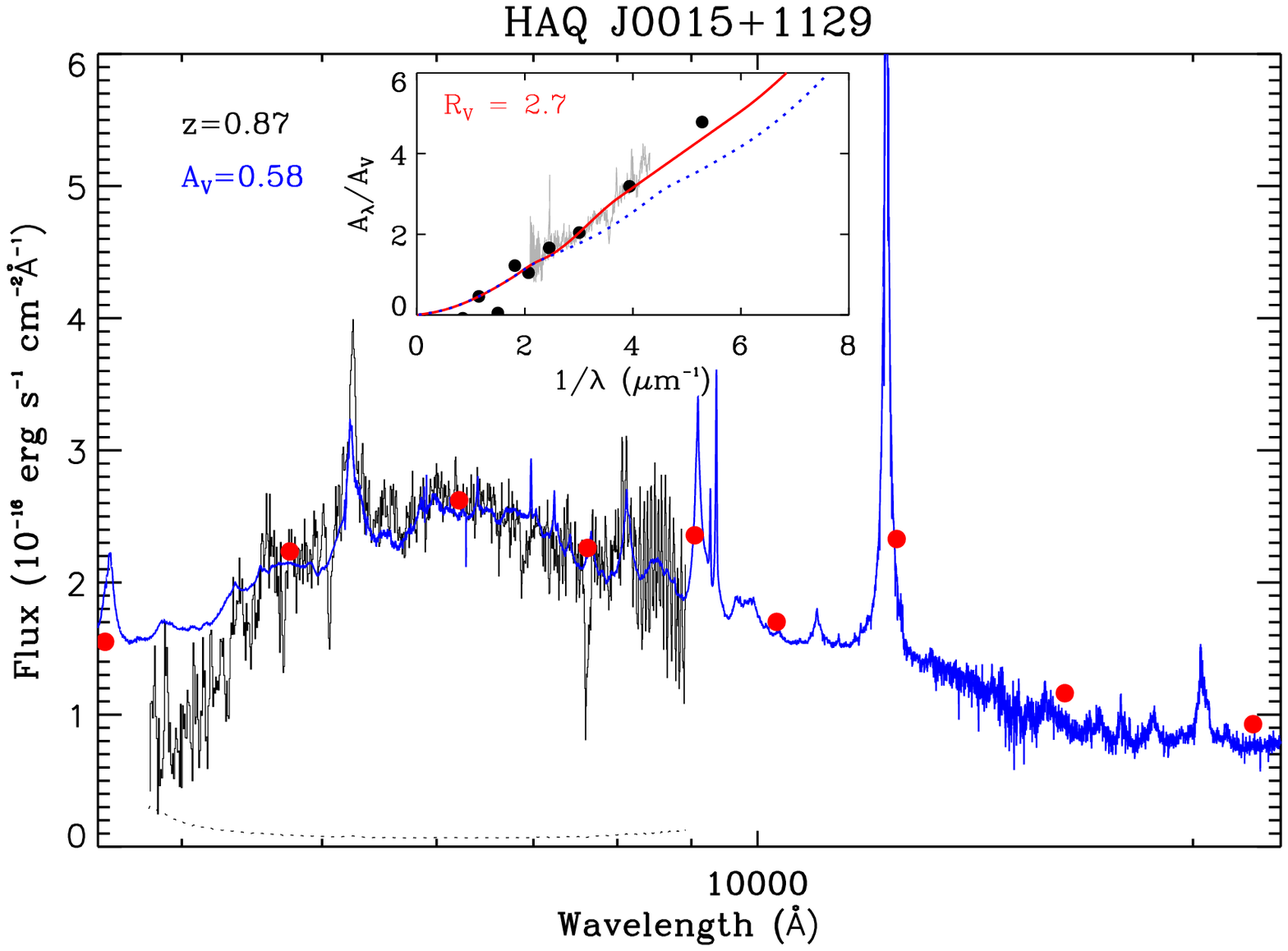}}
{\includegraphics[width=0.63\columnwidth,clip=]{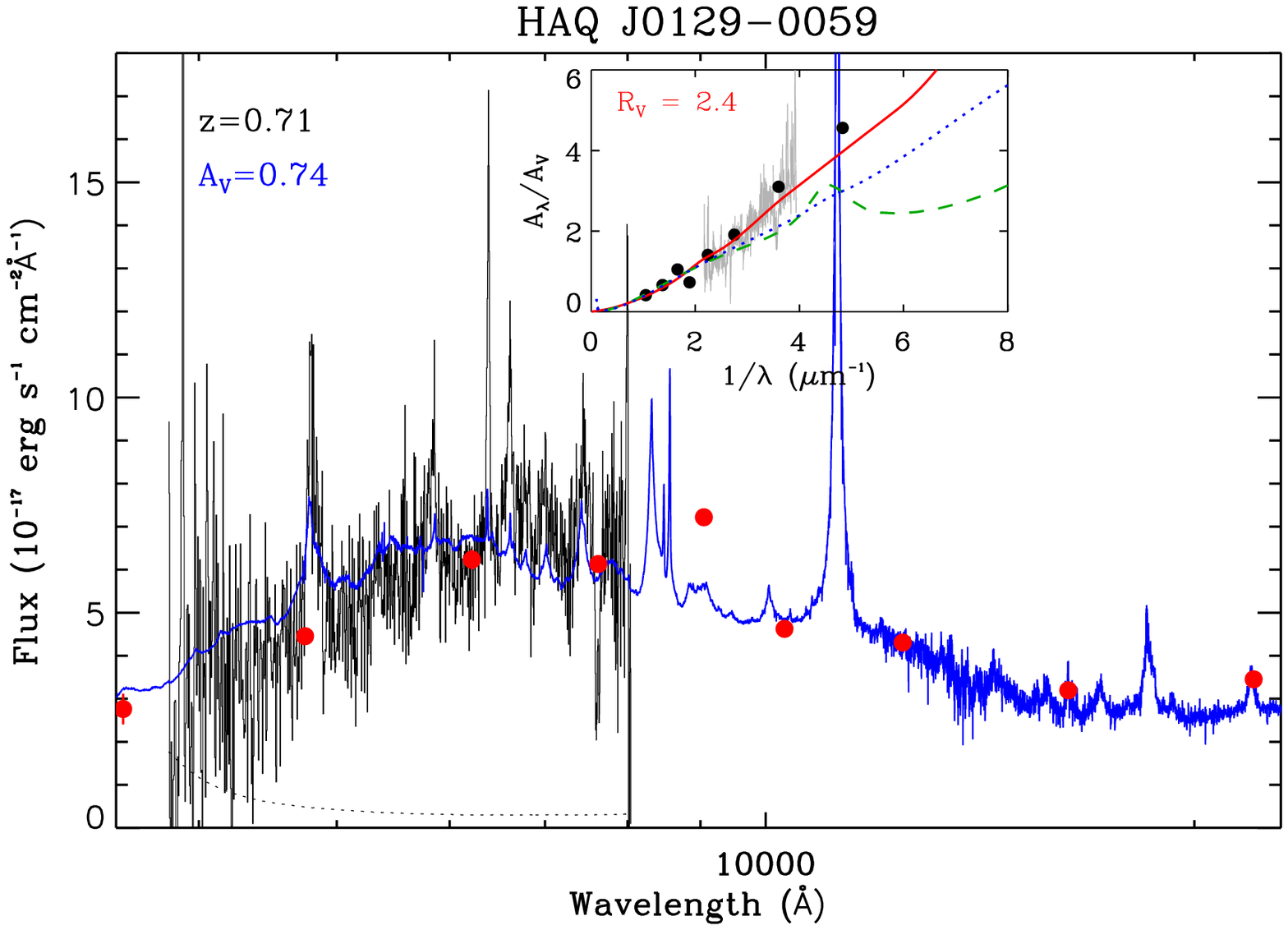}}
{\includegraphics[width=0.63\columnwidth,clip=]{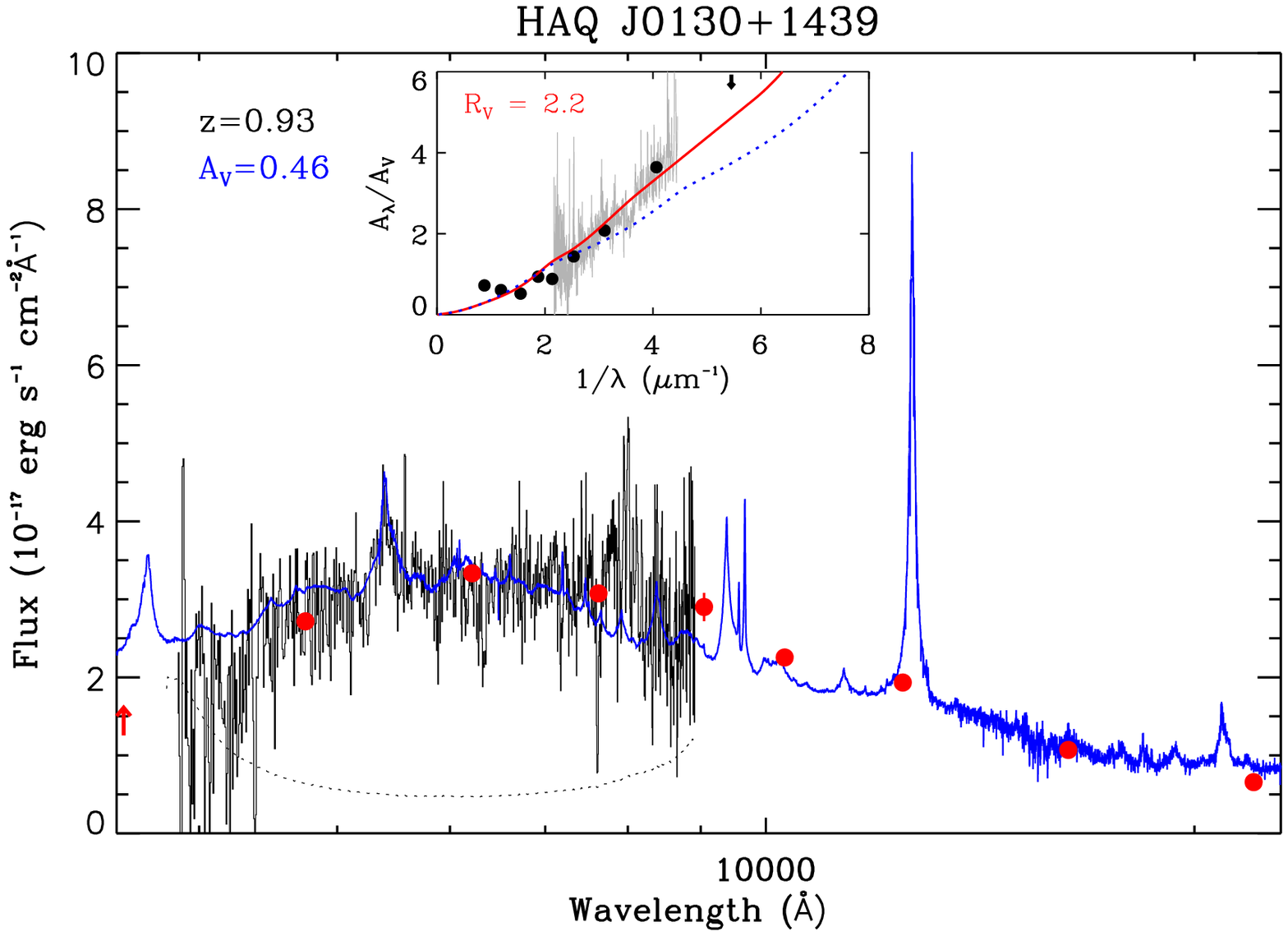}}
{\includegraphics[width=0.63\columnwidth,clip=]{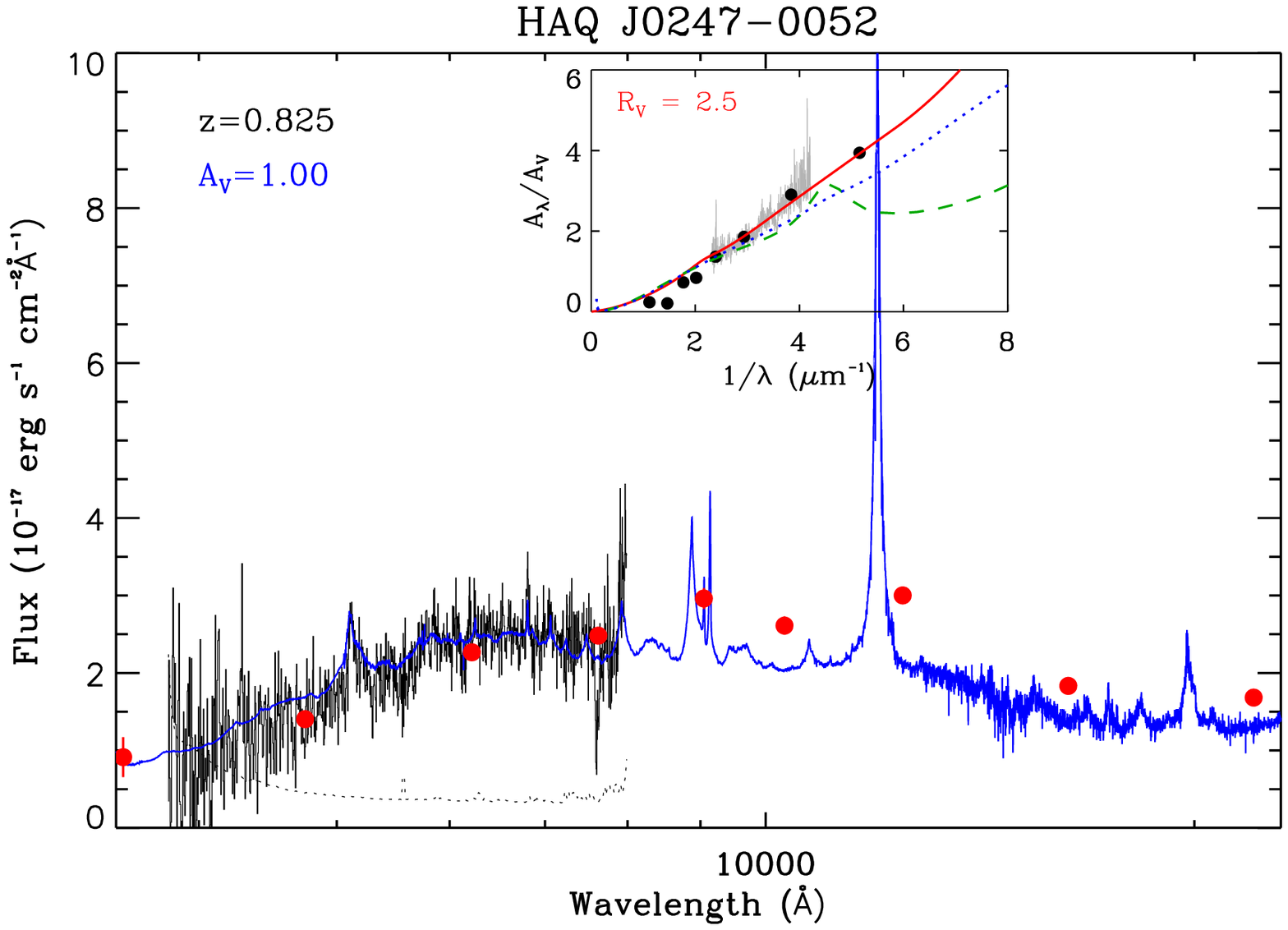}}
{\includegraphics[width=0.63\columnwidth,clip=]{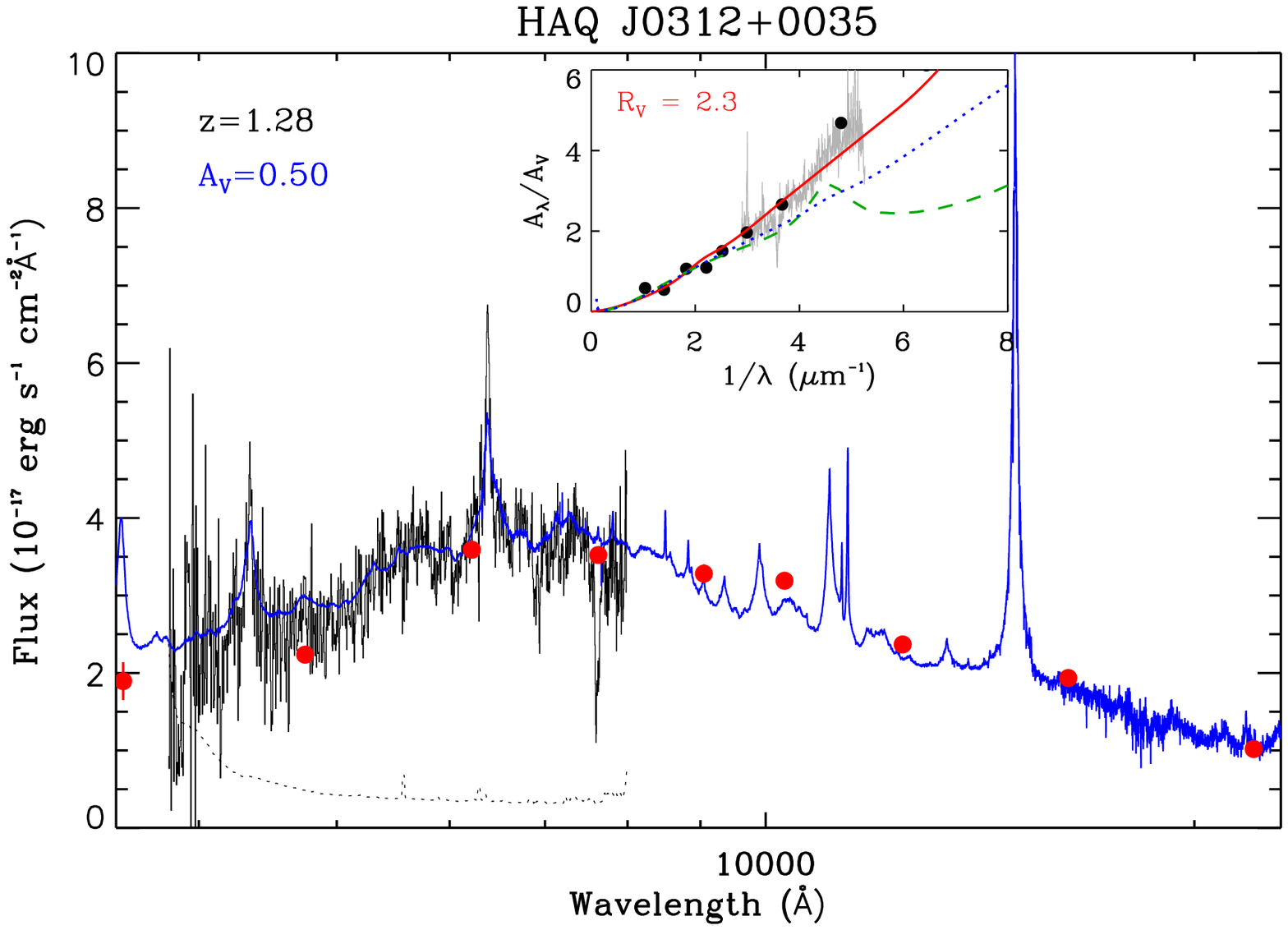}}
{\includegraphics[width=0.63\columnwidth,clip=]{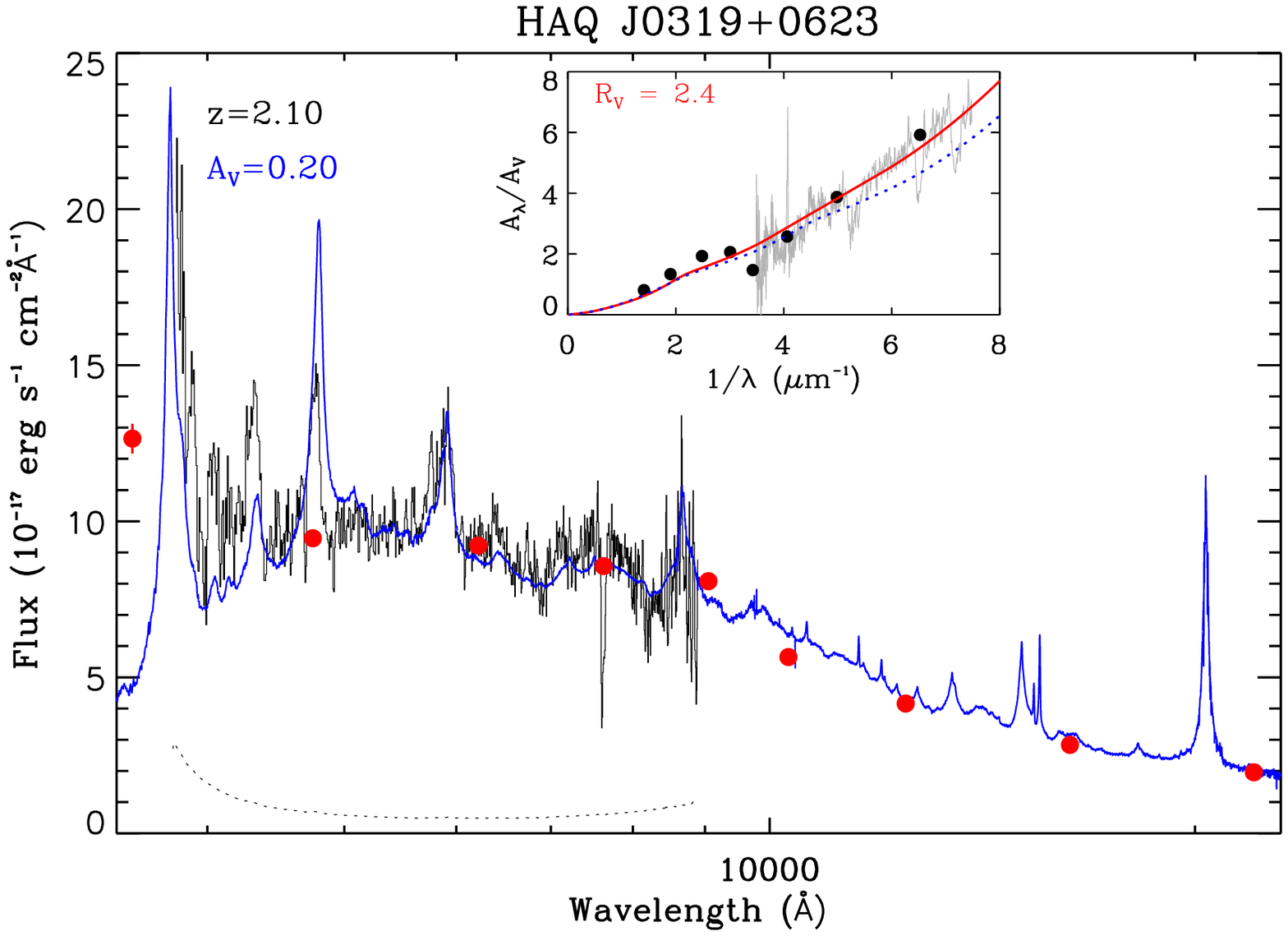}}
{\includegraphics[width=0.63\columnwidth,clip=]{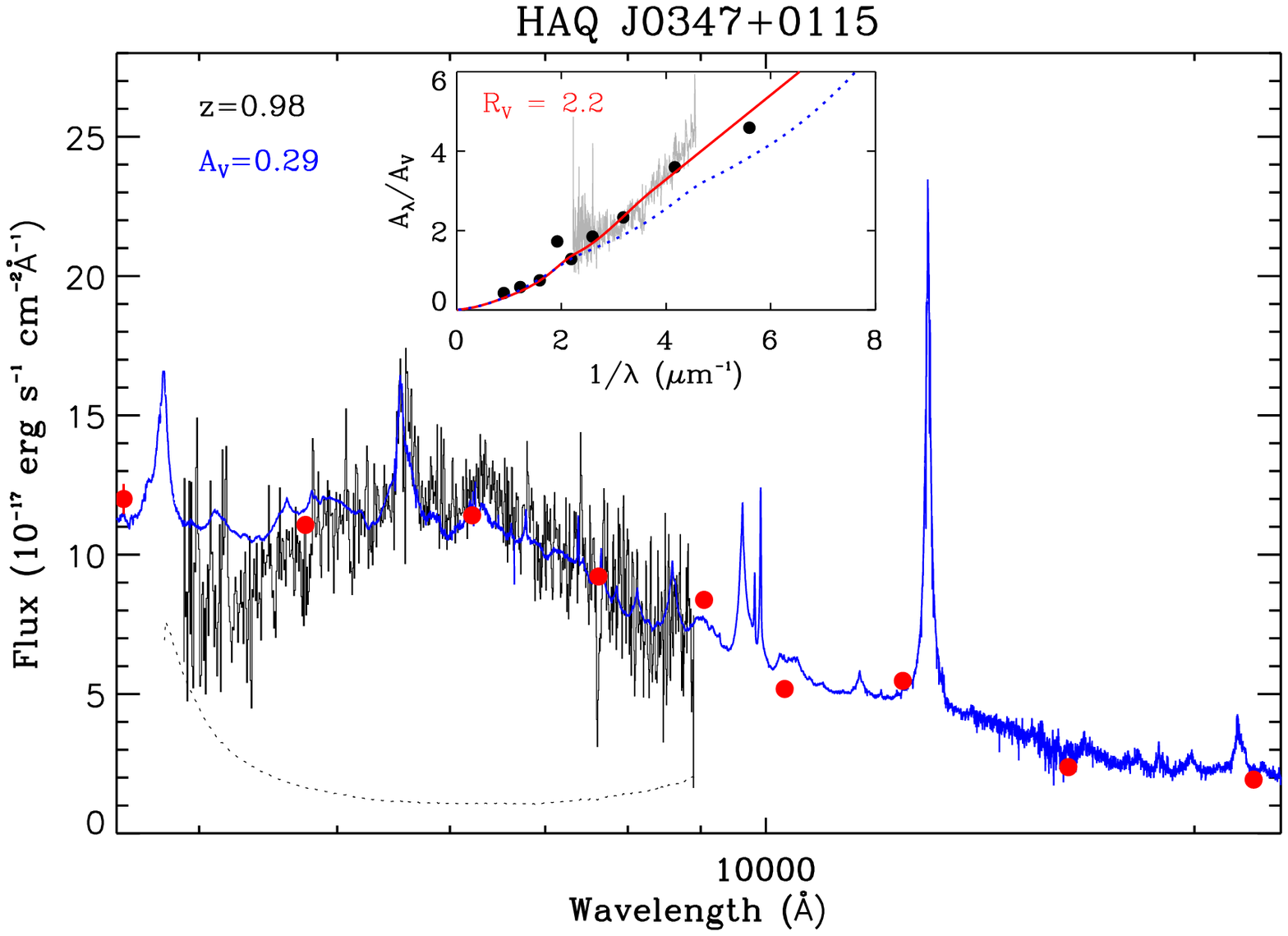}}
{\includegraphics[width=0.63\columnwidth,clip=]{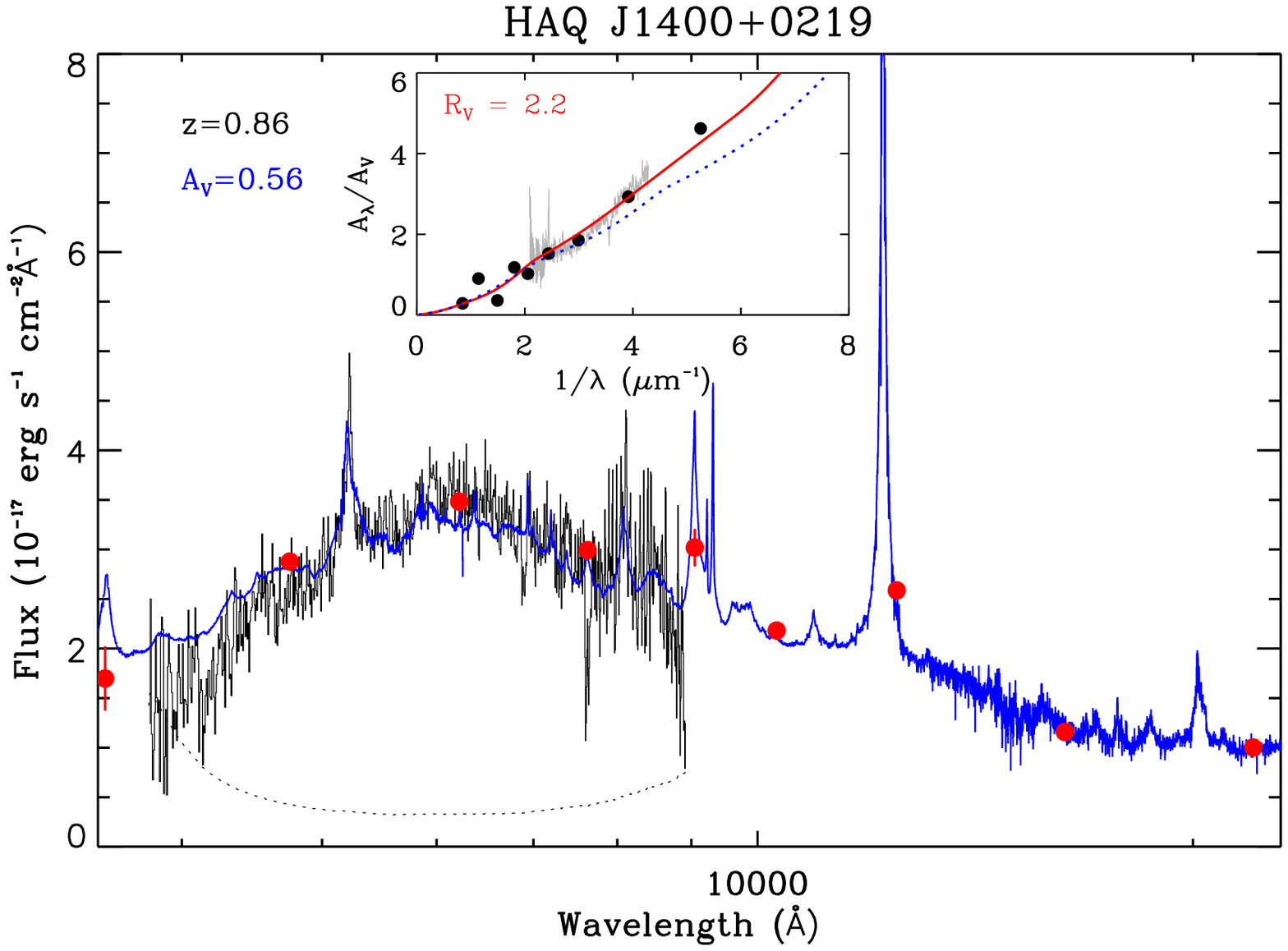}}
{\includegraphics[width=0.63\columnwidth,clip=]{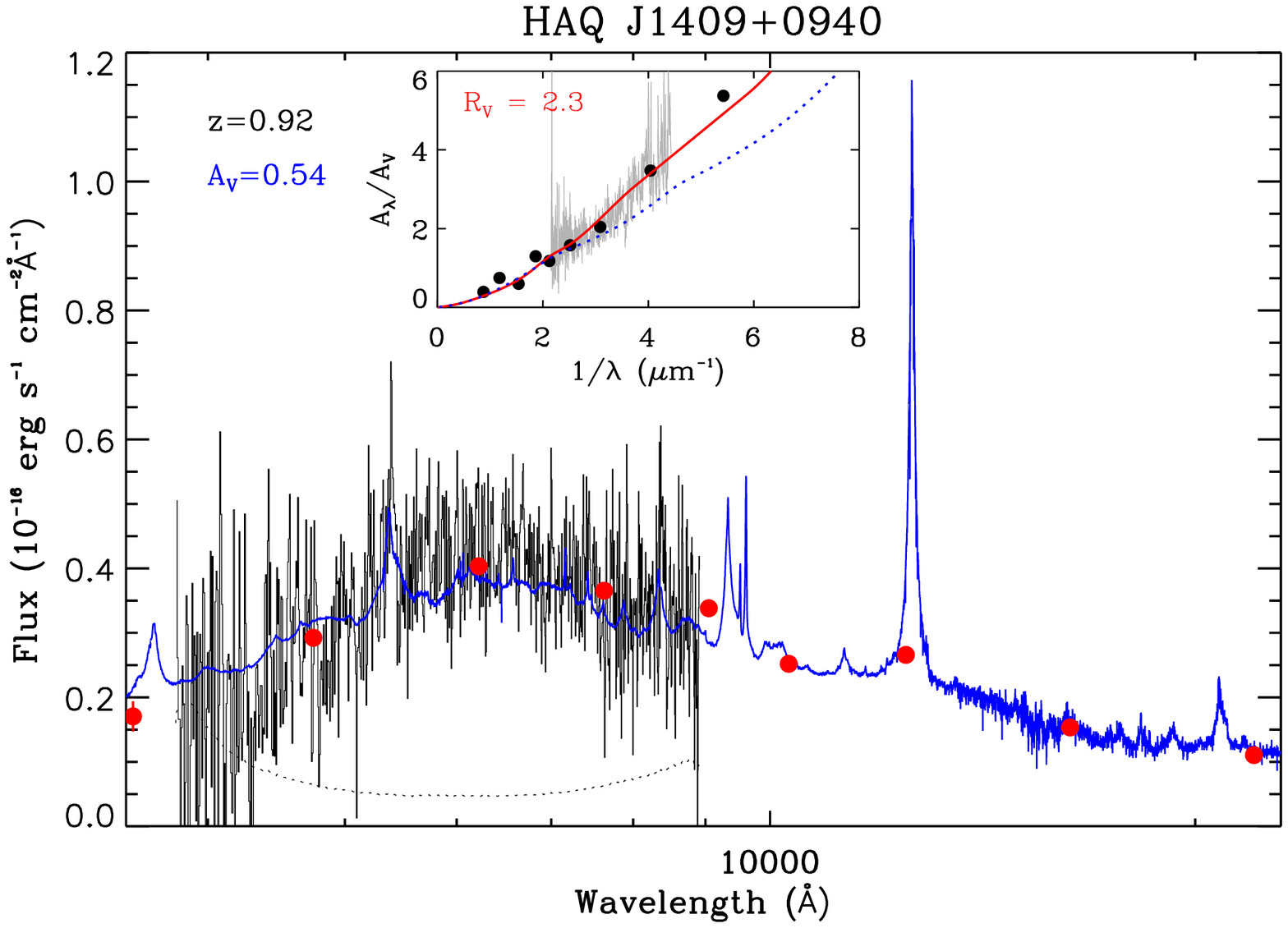}}
{\includegraphics[width=0.63\columnwidth,clip=]{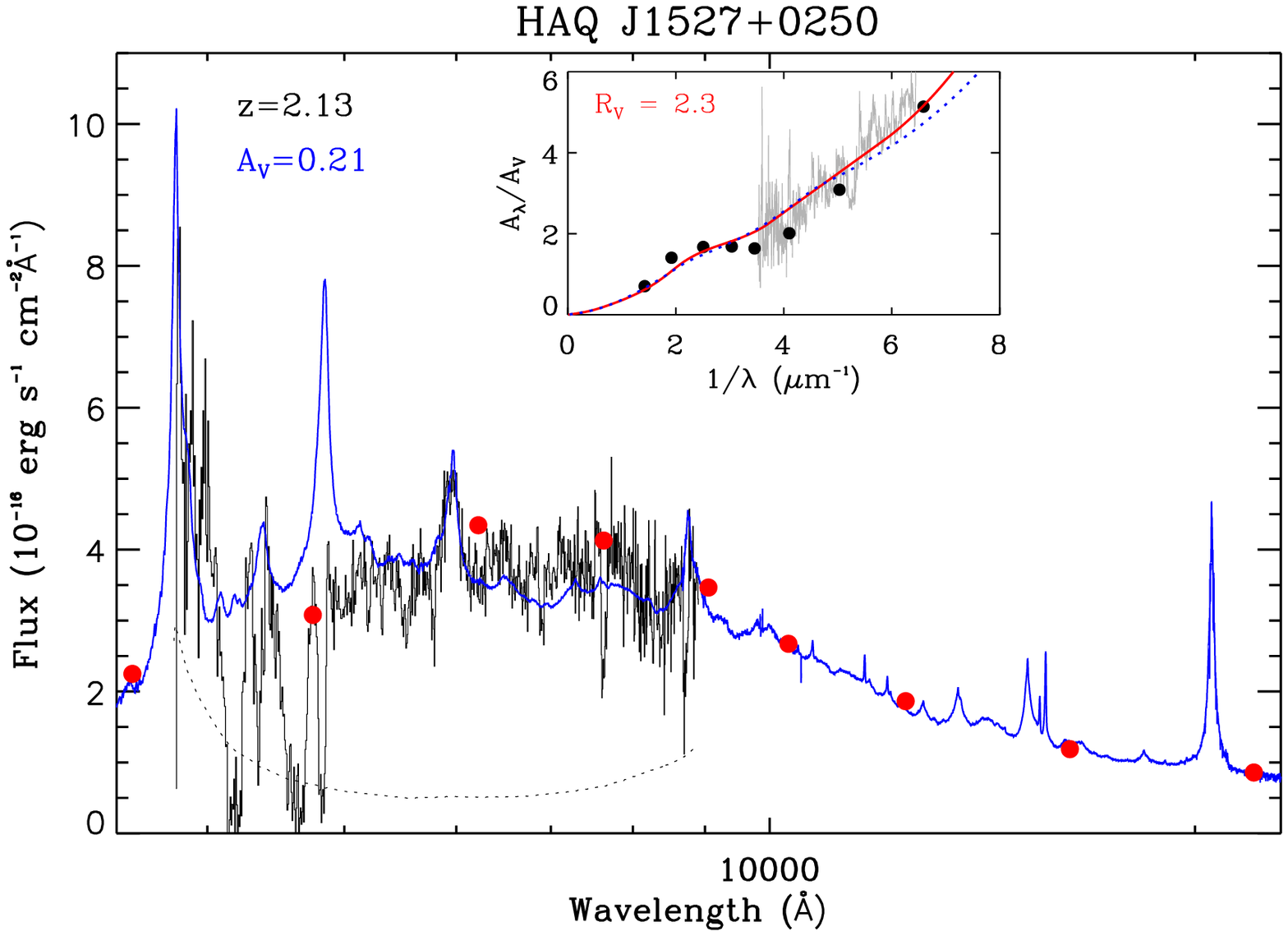}}
{\includegraphics[width=0.63\columnwidth,clip=]{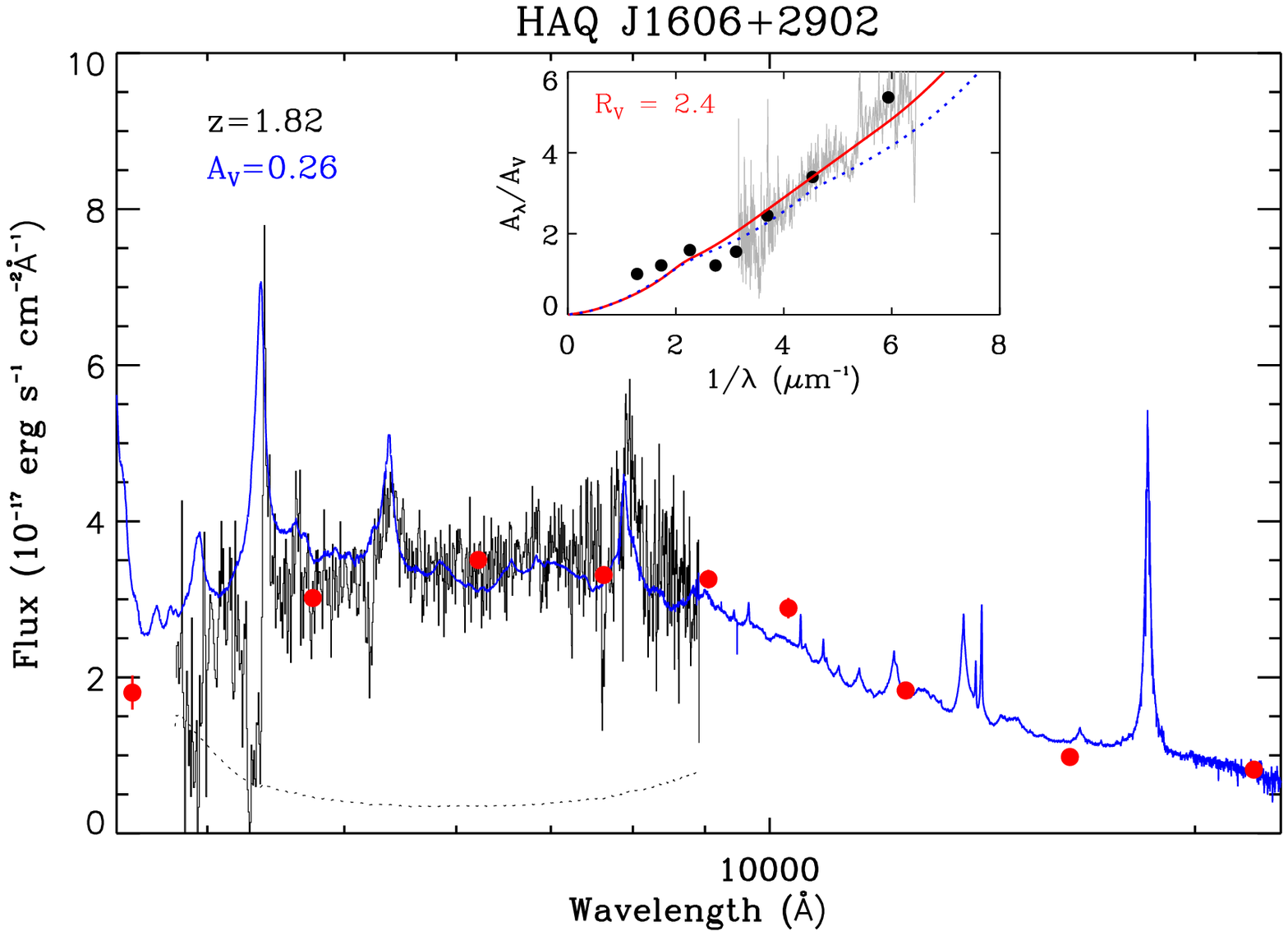}}
{\includegraphics[width=0.63\columnwidth,clip=]{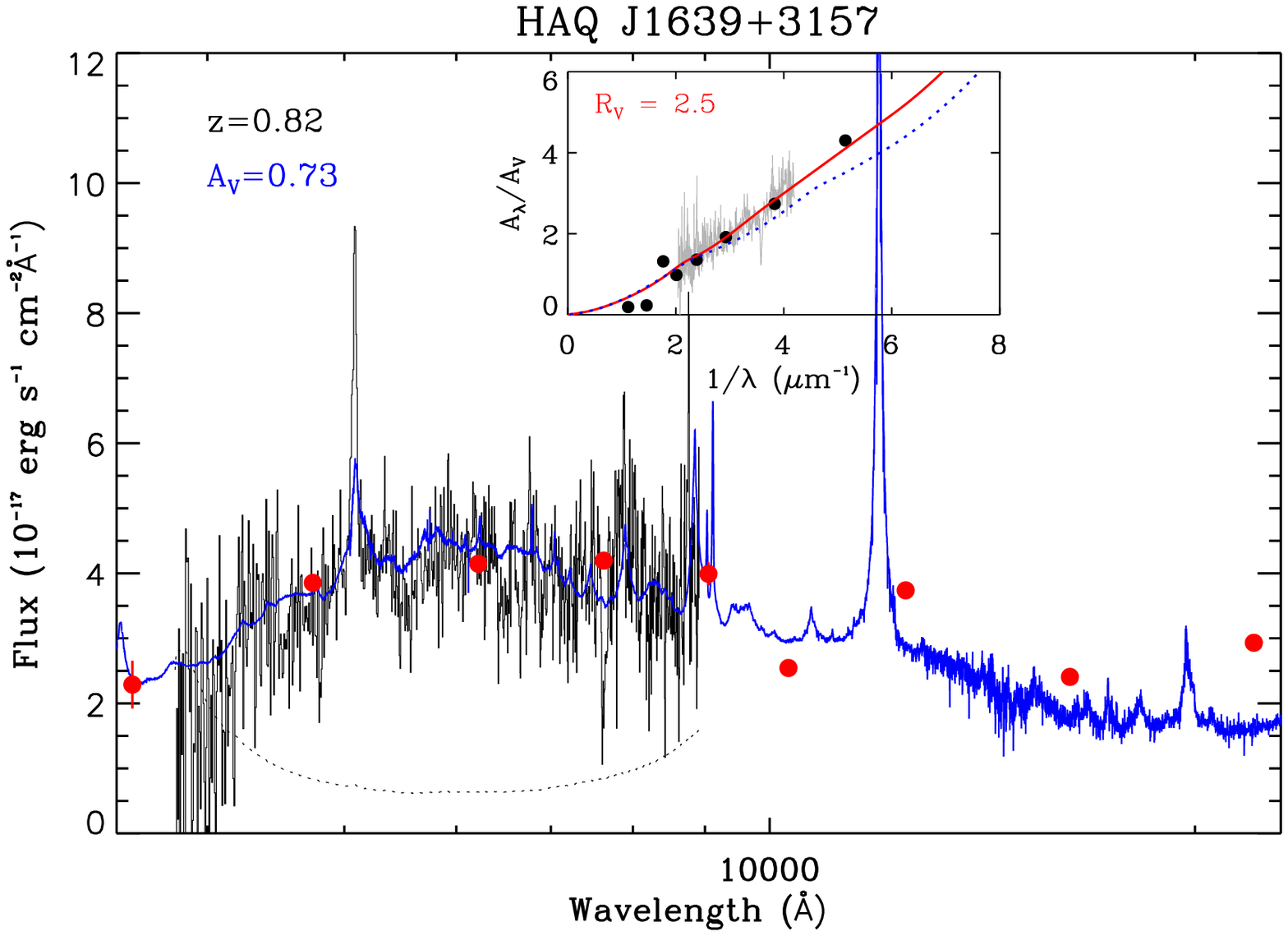}}
{\includegraphics[width=0.63\columnwidth,clip=]{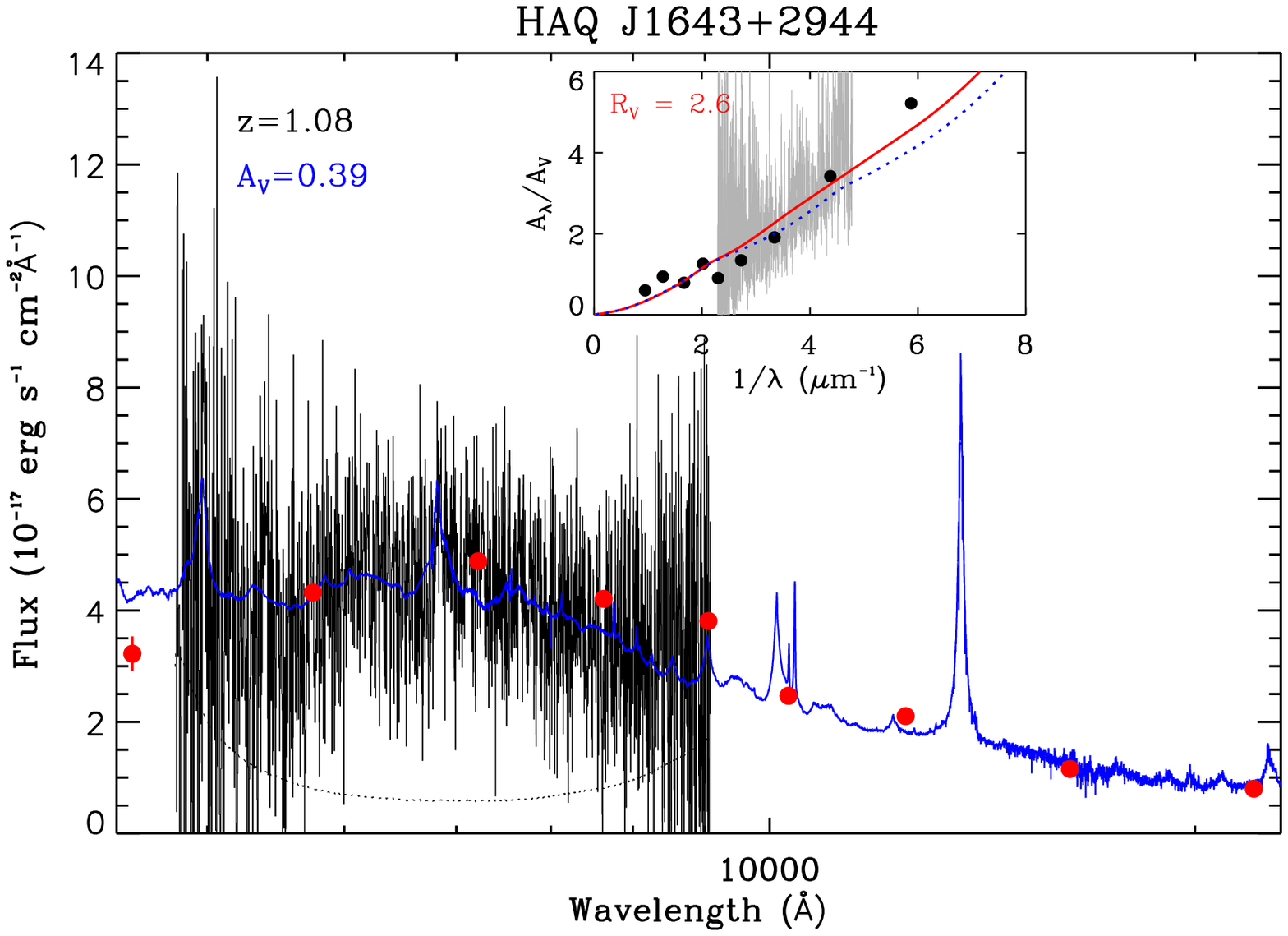}}
{\includegraphics[width=0.63\columnwidth,clip=]{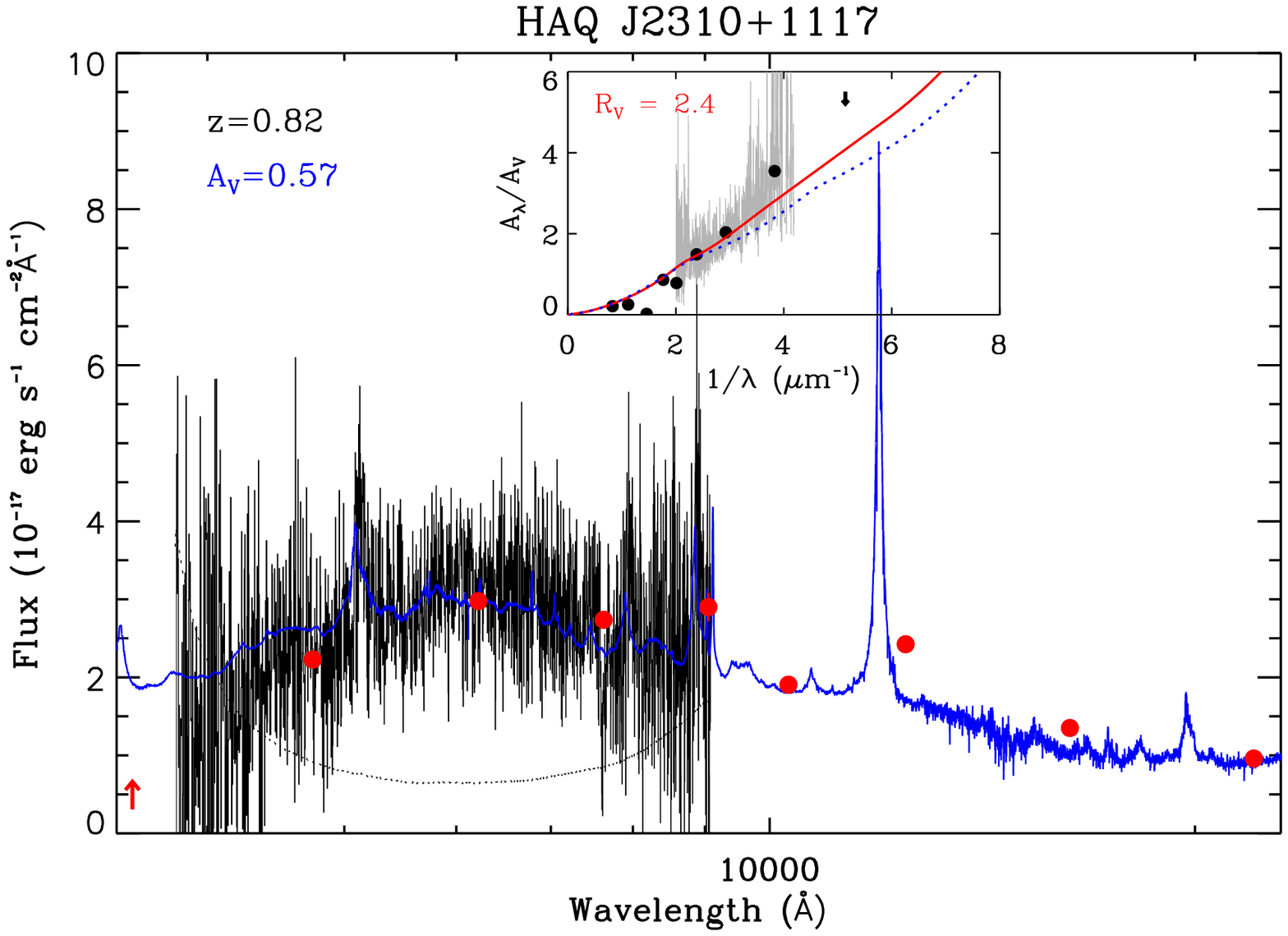}}
{\includegraphics[width=0.63\columnwidth,clip=]{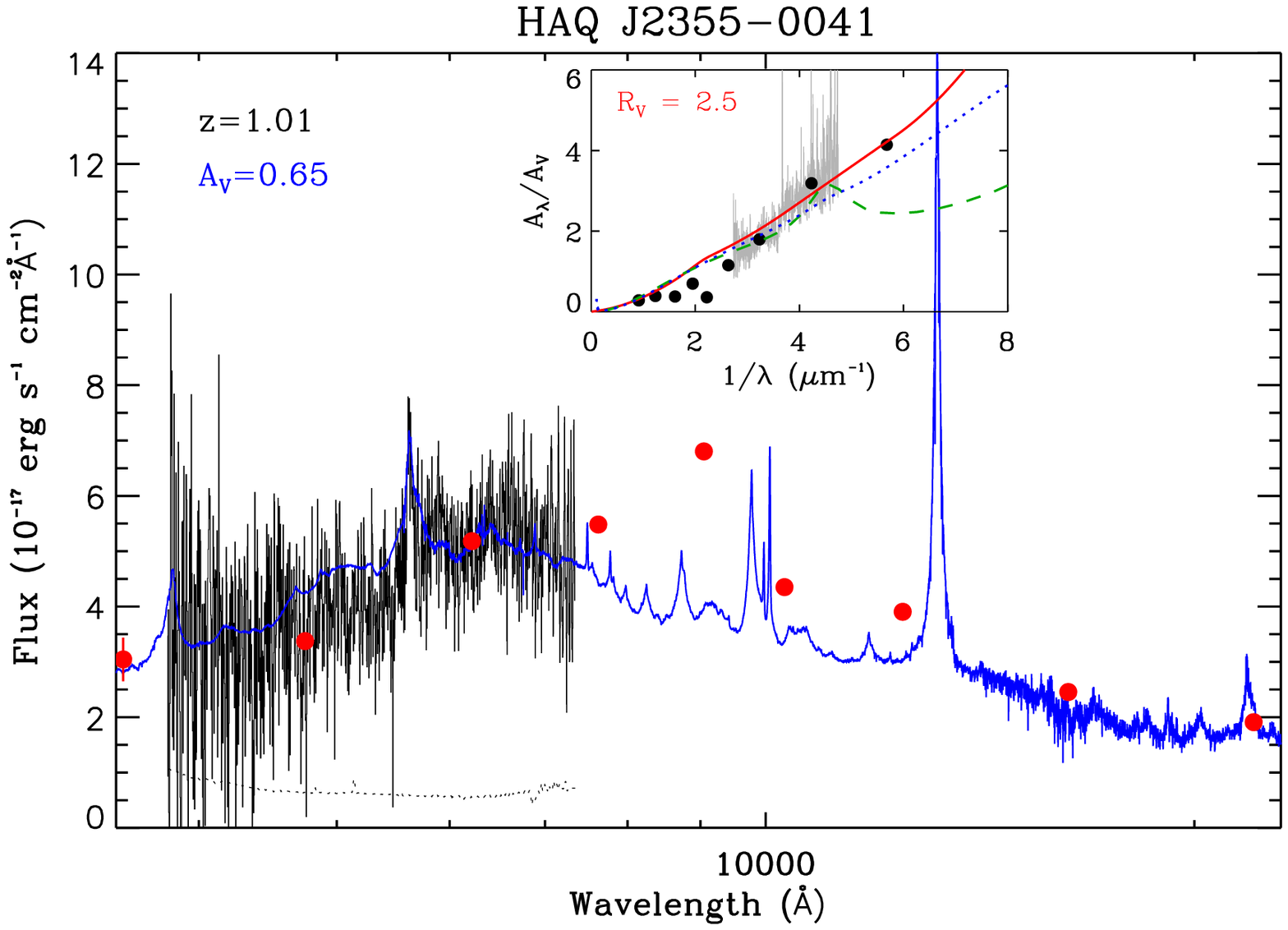}}
     \caption{Same as caption of Fig. \ref{q0151}.}
         \label{qso:fits}
           \end{figure*}

We usually find an extra UV-flux drop in the spectra. However, the $u$-band photometry is consistent with the reddened QSO template. This drop might be caused by the inaccurate flux calibration at the blue end of the spectrum. The QSO SEDs and extinction curves are presented in Fig.~\ref{qso:fits}.
\subsection{HAQ\,J0015$+$1129}
This is a reddened QSO from the HAQ survey of paper\,II at redshift $z=0.87$. The spectrum was obtained with the NOT/ALFOSC. The redshift was based on the detection of the $\ion{Mg}{ii}$ emission line. The data are well fit with a steeper extinction curve with $A_V=0.58\pm0.13$ and $R_V=2.66\pm0.31$. Previously, paper\,II reported $A_V=0.78$.
\subsection{HAQ\,J0129$-$0059}
This is a reddened QSO from the HAQ survey of paper\,I at redshift $z=0.71$. The spectrum was obtained with the NTT/EFOSC2. The redshift was estimated from the $\ion{Mg}{ii}$ emission line. We modelled the QSO SED, and the data are well fit with $A_V=0.74\pm0.17$ and $R_V=2.41\pm0.28$. Previously, paper\,I derived $A_V=1.5$.
\subsection{HAQ\,J0130$+$1439}
This is a reddened QSO from the HAQ survey of paper\,II at redshift $z=0.93$. The spectrum was obtained with the NOT/ALFOSC, and the redshift was derived from the $\ion{Mg}{ii}$ emission line. We modelled the QSO SED, and the data are well fit with $A_V=0.46\pm0.12$ and $R_V=2.23\pm0.35$. Paper\,II reported $A_V=0.4$ from their SED analysis.
\subsection{HAQ\,J0151$+$0618}
This is a reddened QSO from the HAQ survey of paper\,II at redshift $z=0.95$. The spectrum was obtained with the NOT/ALFOSC. The redshift was based on the detection of the $\ion{Mg}{ii}$ emission line. The SED is nicely reproduced with a steep extinction curve with $A_V=0.51\pm0.14$ and $R_V=2.29\pm0.23$. Previously, paper\,II reported $A_V=0.72$.
\subsection{HAQ\,J0247$-$0052}
This is a reddened QSO from the HAQ survey of paper\,I at redshift $z=0.825$. The spectrum was obtained with the NTT/EFOSC2. The redshift was estimated from the $\ion{Mg}{ii}$ and $\ion{O}{ii}$ emission lines. The SED can be modelled with $R_V=2.52\pm0.19$ and a very high extinction value of $A_V=1.00\pm0.11$. Previously, paper\,I inferred $A_V=1.5$ from their SED analysis. 
\subsection{HAQ\,J0312$+$0035}
This is a reddened QSO from the HAQ survey of paper\,I at redshift $z=1.28$. The spectrum was obtained with the NTT/EFOSC2. The redshift was derived from the $\ion{Mg}{ii}$ and $\ion{C}{iii}$ emission lines. The SED can be modelled with a steep extinction curve with $A_V=0.50\pm0.12$ and $R_V=2.29\pm0.23$. Previously, paper\,I derived $A_V=0.8$.
\subsection{HAQ\,J0319$+$0623}
This is a reddened QSO from the HAQ survey of paper\,II at redshift $z=2.10$. The spectrum was obtained with the NOT/ALFOSC, and the redshift was derived from the $\ion{Si}{iv}+\ion{O}{iv}$, $\ion{C}{iv}$, and $\ion{C}{iii}$ emission lines. The SED can be modelled with a steep extinction curve with $A_V=0.20\pm0.06$ and $R_V=2.38\pm0.365$. Paper\,II reported $A_V=0.13$ from the SED analysis.
\subsection{HAQ\,J0347$+$0115}
This is a reddened QSO from the HAQ survey of paper\,II at redshift $z=0.98$. The spectrum was obtained with the NOT/ALFOSC. The redshift was estimated from the $\ion{Mg}{ii}$ emission line. The SED is modelled well with a steep extinction curve and $A_V=0.29\pm0.12$ together with $R_V=2.23\pm0.39$. Previously, paper\,II inferred $A_V=0.4$ from the SED analysis.
\subsection{HAQ\,J1400$+$0219}
This is a reddened QSO from the HAQ survey of paper\,II at redshift $z=0.86$. The spectrum was obtained with the NOT/ALFOSC. The redshift was based on the detection of $\ion{Mg}{ii}$ emission line. We found the observed data fit very well with a steep extinction curve and relatively high $A_V$ with $A_V=0.56\pm0.09$ and $R_V=2.16\pm0.36$. Previously, paper\,II reported $A_V=0.85$. 
\subsection{HAQ\,J1409$+$0940}
This is a reddened QSO from the HAQ survey of paper\,II at redshift $z=0.92$. The spectrum was obtained with the NOT/ALFOSC, and the redshift was derived from the relatively weak $\ion{Mg}{ii}$ emission line. The SED can be modelled with a steep extinction curve with $A_V=0.54\pm0.10$ and $R_V=2.27\pm0.26$. Paper\,II derived $A_V=0.93$ from SED analysis. 
\subsection{HAQ\,J1527$+$0250}
This is a reddened BAL QSO from the HAQ survey of paper\,II at redshift $z=2.13$. The spectrum was obtained with the NOT/ALFOSC. The redshift was estimated from the $\ion{C}{iv}$ and $\ion{C}{iii}$ emission lines. We found that the observed data fit very well with a steep extinction curve with $A_V=0.21\pm0.07$ and $R_V=2.33\pm0.36$. Previously, paper\,II reported $A_V=0.2$.
\subsection{HAQ\,J1606$+$2902}
This is a reddened BAL QSO from the HAQ survey of paper\,II at redshift $z=1.82$. The spectrum was obtained with the NOT/ALFOSC, and the redshift was based on the detection of $\ion{C}{iv}$, $\ion{Fe}{ii}$, and $\ion{Mg}{ii}$ emission lines. The data are nicely reproduced with a low extinction value of $A_V=0.26\pm0.08$ and $R_V=2.44\pm0.24$. Paper\,II inferred $A_V=0.25$ from their SED analysis.
\subsection{HAQ\,J1639$+$3157}
This is a reddened QSO from the HAQ survey of paper\,II at redshift $z=0.82$. The spectrum was obtained with the NOT/ALFOSC, and the redshift was based on the detection of $\ion{Mg}{ii}$ emission line. The data are well described with a steep extinction curve and relatively high extinction value of $A_V=0.73\pm0.11$ and $R_V=2.48\pm0.18$. Previously, paper\,II derived $A_V=0.79$.
\subsection{HAQ\,J1643$+$2944}
This is a reddened QSO from the HAQ survey of paper\,II at redshift $z=1.08$. The spectrum was obtained with the NOT/ALFOSC. The redshift was derived from the relatively weak $\ion{Mg}{ii}$ emission line. The SED of the QSO fits well with $A_V=0.39\pm0.13$ and $R_V=2.56\pm0.27$. Paper\,II previously reported $A_V=0.44$ from the SED analysis.
\subsection{HAQ\,J2310$+$1117}
This is a reddened QSO from the HAQ survey of paper\,II at redshift $z=0.82$. The spectrum was obtained with the NOT/ALFOSC, and the redshift was estimated from the weak $\ion{Mg}{ii}$ emission line. We found that the observed data fit very well with a steep extinction curve with $A_V=0.57\pm0.13$ and $R_V=2.44\pm0.28$. Previously, paper\,II reported $A_V=0.92$ from the SED analysis.
\subsection{HAQ\,J2355$-$0041}
This is a reddened QSO from the HAQ survey of paper\,I at redshift $z=1.01$. The spectrum was obtained with the NOT/ALFOSC. The redshift was inferred from the $\ion{Mg}{ii}$ emission line. The SED is fit very well with a steep extinction curve with $A_V=0.65\pm0.13$ and $R_V=2.54\pm0.24$. Paper\,I derived $A_V=1.0-1.6$ from the SED analysis.
\end{appendix}

\begin{appendix}
\end{appendix}

\end{document}